\documentclass[conference]{IEEEtran}
\IEEEoverridecommandlockouts

\usepackage{graphicx}  %Required
\usepackage{commath}
\usepackage{verbatim}
\usepackage{multirow}
\usepackage{float} 

\usepackage{cite}
\usepackage{amsmath,amssymb,amsfonts}
\usepackage{algorithmic}
\usepackage{textcomp}
\usepackage{xcolor}

\usepackage{setspace}

\usepackage[font={footnotesize}]{subcaption}
\newcommand{\subparagraph}{}
 \usepackage{titlesec}
 \titlespacing\section{0pt}{6pt plus 2pt minus 1pt}{4pt plus 2pt minus 1.5pt}
 \titlespacing\subsection{0pt}{4pt plus 2pt minus 1pt}{2pt plus 2pt minus 1pt}
 \titlespacing\subsubsection{10pt}{3pt plus 0pt minus 2pt}{2pt plus 0pt minus 2pt}
\titleformat{\subsubsection}[runin]{\normalfont\normalsize\itshape}{\arabic{subsubsection})}{5pt}{}[:\,\,]

\usepackage{cancel}

\begin{document}
\title{Online Collection and Forecasting of Resource Utilization in Large-Scale Distributed Systems \vspace{-0.1in}}

\author{
\IEEEauthorblockN{Tiffany Tuor\IEEEauthorrefmark{1}, Shiqiang Wang\IEEEauthorrefmark{2}, Kin K. Leung\IEEEauthorrefmark{1}, Bong Jun Ko\IEEEauthorrefmark{2}}
\IEEEauthorblockA{\IEEEauthorrefmark{1}Imperial College London, UK. Email: \{tiffany.tuor14, kin.leung\}@imperial.ac.uk}
\IEEEauthorblockA{\IEEEauthorrefmark{2}IBM T. J. Watson Research Center, Yorktown Heights, NY, USA. Email: \{wangshiq, bko\}@us.ibm.com}
\vspace{-0.3in}
\thanks{This research was sponsored by the U.S. Army Research Laboratory and the U.K. Ministry of Defence under Agreement Number W911NF-16-3-0001. The views and conclusions contained in this document are those of the authors and should not be interpreted as representing the official policies, either expressed or implied, of the U.S. Army Research Laboratory, the U.S. Government, the U.K. Ministry of Defence or the U.K. Government. The U.S. and U.K. Governments are authorized to reproduce and distribute reprints for Government purposes notwithstanding any copyright notation hereon.}
}

\maketitle
\begin{abstract}
Large-scale distributed computing systems often contain thousands of distributed nodes (machines). Monitoring the conditions of these nodes is important for system management purposes, which, however, can be extremely resource demanding as this requires collecting local measurements of each individual node and constantly sending those measurements to a central controller. Meanwhile, it is often useful to forecast the future system conditions for various purposes such as resource planning/allocation and anomaly detection, but it is usually too resource-consuming to have one forecasting model running for each node, which may also neglect correlations in observed metrics across different nodes. In this paper, we propose a mechanism for collecting and forecasting the resource utilization of machines in a distributed computing system in a scalable manner. We present an algorithm that allows each local node to decide when to transmit its most recent measurement to the central node, so that the transmission frequency is kept below a given constraint value. Based on the measurements received from local nodes, the central node summarizes the received data into a small number of clusters. Since the cluster partitioning can change over time, we also present a method to capture the evolution of clusters and their centroids. As an effective way to reduce the amount of computation, time-series forecasting models are trained on the time-varying centroids of each cluster, to forecast the future resource utilizations of a group of local nodes. The effectiveness of our proposed approach is confirmed by extensive experiments using multiple real-world datasets.
\end{abstract}

\section{Introduction}

Modern cloud computing systems often include thousands of machines that process tasks originating from different geographical regions. The effective management of such large-scale distributed systems is very challenging. For example, even within a single data center, where machines are interconnected with high-speed networking and often owned by a single service provider, it is very difficult to allocate resources optimally, due to the high variation of resource demands of different tasks. As a result, resource over-provisioning (allocating too much resource) and under-provisioning (allocating too little resource) often occur in practical cloud systems~\cite{CaiLessProvisioning2018,Grechanik2016}. The former causes waste in resources and high operational cost, and the latter causes degradation in user experience.

To overcome these issues, we need to precisely monitor and predict the resource utilization (such as CPU and memory utilization) of individual machines~\cite{silvestri2015online}, based on which the current and future available resources at each machine can be inferred so that the system can be properly managed and resource allocation can be performed in a near-optimal way. 
In particular, measurements of the resource utilization at each physical machine (\emph{local node}) has to be transmitted to a central controller (\emph{central node}). The controller needs to forecast the future resource availability of each machine, so that it can assign new incoming tasks to machines that are predicted to have the most suitable amount of available resources.
Furthermore, the forecasting has to be done in an \emph{online} manner, where algorithms make decisions based on information received up to the current time, and does not assume knowledge of future information due to obvious practical reasons.

There exist several challenges towards a distributed system that can perform the above functionalities of collecting and forecasting resource utilization. 
First, it is often bandwidth-consuming and unnecessary to transmit all the measurements collected at local nodes to the central node. Second, predictive models for data forecasting typically have high complexity, thus running a forecasting model for the time-series measurement data collected at each local node would consume too much computational resource. Third, measurements at each local node are collected in an online manner, which form a time series; decisions related to data collection and forecasting need to be made in an online manner as well.

In this paper, we address the above challenges and propose a mechanism that efficiently collects and forecasts the resource utilization at machines in a large-scale distributed system. The results provided by our mechanism can be used for system management such as resource allocation. We focus on the collection and forecasting of resource utilization in this paper, and leave its application to system management for future work. Our main contributions are summarized as follows.
\begin{enumerate}
    \item We propose an algorithm for each local node to adaptively decide when to transmit its latest measurement to the central node, subject to a maximum frequency of transmissions that is given as a system-constraint parameter. The algorithm adapts to the degree of changes in observations since the last transmission, so that the allowed transmission bandwidth is efficiently used.
    \item We propose a dynamic clustering algorithm for the central node to partition the measurements received from local nodes into a given number of clusters. The algorithm allows the clustering to evolve over time, and the cluster centroids are a compressed representation of the dynamic observations of the large distributed system. 
    \item We propose a forecasting mechanism where the centroids of each cluster evolving over time constitute a time series that is used to train a forecasting model. The trained model is then used to forecast the future resource utilizations of a group of local nodes. 
    \item Extensive experiments of our proposed mechanism have been conducted using three real-world computing cluster datasets, to show the effectiveness of our proposed approach.
\end{enumerate}

The clustering, model training, and forecasting are all performed in an online manner, based on ``intermittent'' measurement data received at the central node. 

The rest of this paper is organized as follows. In the next section, we review the related work. In Section~\ref{section:problem-formulation}, we present the system overview together with some definitions. The proposed algorithms are described in Section~\ref{section:proposed-framework}. The experimentation settings and results are given in Section~\ref{section: experiments}, and Section~\ref{section:conclusion} draws our conclusion.

\section{Related Work}\label{section:related}

The existing body of work that uses prediction/forecasting models to assist resource scheduling mostly focuses on aggregated workloads or resource demands that can be described as a single time series~\cite{CaiLessProvisioning2018,Grechanik2016,Nikravesh2015,Shen2018}. While these approaches are useful for predicting the future demand, they do not capture the dynamics of resource utilization at individual physical machines, and hence cannot predict how much resource is utilized or available in the physical system. In this paper, we focus on resource utilization at machines in the distributed system, which is more complex because each machine generates a time-series measurement data on its own.

Some existing approaches for efficient data collection in a distributed system involve only a selected subset of local nodes that transmit data to the central node~\cite{coluccia2011lossy,barcelo2012enhanced,anagnostopoulos2014advanced,li2018compressed,leinonen2014compressed,zhang2016near,krause2008near,silvestri2015online}. 
More specifically, techniques in \cite{zhang2016near,krause2008near,silvestri2015online} select the best set of monitors (local nodes) subject to a constraint on the number of monitors, and infer data from the unobserved local nodes based on Gaussian models. 
Methods in \cite{coluccia2011lossy,barcelo2012enhanced,anagnostopoulos2014advanced,li2018compressed,leinonen2014compressed} are based on compressed sensing, where a subset of local nodes is randomly selected to collect data, then matrix completion is applied to reconstruct data from unobserved nodes. The approaches based on compressed sensing generally perform worse than Gaussian-based approaches~\cite{silvestri2015online}.
All these approaches where only some of the local nodes send data in the monitoring phase lead to unbalanced resource consumption (such as communication bandwidth, energy, etc.).

To avoid unbalanced resource consumption, some existing approaches
consider settings where every node sends data to the central node but with a sampling rate adapted directly at each node \cite{liu2005energy,law2009energy,harb2016adaptive,chatterjea2008adaptive,idrees2017distributed}. 
However, the sampling rate in these works is only implicitly related to the transmission frequency. None of them allows one to specify a target transmission frequency which is proportional to the required communication bandwidth. In this paper, we propose an algorithm that decides when to transmit subject to a maximum transmission frequency. This allows the system to explicitly specify the communication budget.

For the clustering of local node measurements, Gaussian models are widely used, such as in~\cite{zhang2016near,krause2008near,silvestri2015online}. However, these methods require a separate training phase to estimate the covariance matrix, during which it needs to collect all the data from all local nodes, which can be bandwidth consuming. In addition, a sufficiently large number of samples are required for a good estimation of the covariance matrix. When the correlation among local nodes vary frequently, which is the case with resource utilization at machines in distributed systems (see Section~\ref{section:motivation-example} for further discussion), the system may not be able to collect enough samples to estimate the covariance matrix with a reasonable accuracy. In this paper, we propose a clustering mechanism that works well with highly varying resource utilization data.

The evolution of clusters over time is related to the area of evolutionary clustering~\cite{chakrabarti2006evolutionary,xu2014adaptive,greene2010tracking,yang2011detecting}, for which typical applications include community matching in social science~\cite{greene2010tracking}, disease diagnosis in bio-informatics~\cite{ma2006evolutionary}, user preference modelling in dynamic recommender systems~\cite{rana2014evolutionary}, etc. To our knowledge, evolutionary clustering techniques have not been applied to the dynamic clustering and forecasting of resource utilization at multiple machines, where the objectives are different from the above applications.

In summary, while there exist methods in the literature that are related to specific parts of our problem, they focus on different scenarios or applications and do not directly apply to our problem, as explained above. Furthermore, to our knowledge, a system/mechanism that efficiently collects and forecasts resource utilization of all machines in a distributed system does not exist in the literature. 
This paper overcomes the challenge of developing such a mechanism with different components working smoothly together, while providing good performance in practical settings.

\section{Motivational Experiment}\label{section:motivation-example}

To illustrate the challenge in the problem we study in this paper, we start with a motivational experiment comparing the long-term spatial correlations\footnote{The (spatial) correlation of two nodes is defined as the sample covariance of measurements obtained at the two nodes divided by the standard deviations of both sets of measurements (each obtained at one of the two nodes)~\cite{Shen2018}.} in resource utilizations at different machines in a distributed computing environment and sensor measurements at different nodes in a sensor network. 

\begin{figure}
    \centering
    
    \begin{subfigure}[b]{0.5\columnwidth}
        \centering
        \includegraphics[width=1\linewidth]{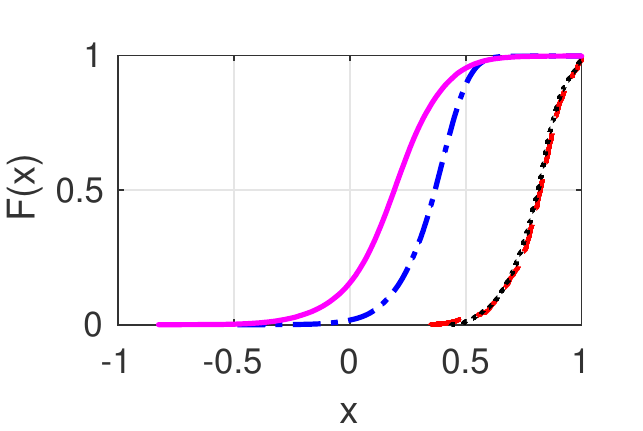}
    \end{subfigure}%
    ~
    \begin{subfigure}[b]{0.5\columnwidth}
        \centering
        \includegraphics[width=0.6\linewidth]{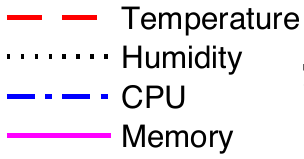}
        \vspace{0.6in}
    \end{subfigure}%    
    \caption{Empirical cumulative distribution function (CDF) of correlation values of different datasets.}
    \label{cdf}
\end{figure}

We consider the sensor network dataset collected by Intel Research Laboratory at Berkeley~\cite{bodik2004intel}, which includes sensor measurements over $12$ days, and the Google cluster usage trace (version 2)~\cite{reiss2011google}, which includes resource utilizations at machines over one month. The empirical cumulative distribution function (CDF) of the spatial correlation values computed on the temperature and humidity data from the sensor dataset and the CPU and memory utilization data (aggregated for all tasks running on each machine) from the Google cluster dataset are plotted in Fig.~\ref{cdf}, where each type of data is considered separately.

We see that for CPU and memory utilization, most of the spatial correlation values are between $-0.5$ and $0.5$, whereas most correlation values are above $0.5$ for temperature and humidity data. This shows that in the long term (over the entire duration of the dataset), the spatial correlation in resource utilization among machines in a distributed computing system is much weaker than the spatial correlation in sensor measurements at different nodes in a sensor network.
Therefore, we \emph{do not} have strong long-term spatial correlation in our scenario, which is required by Gaussian-based methods for covariance matrix estimation (see also the related discussions in Section~\ref{section:related}). Hence, Gaussian models which are widely used in the clustering of sensor network data~\cite{zhang2016near,krause2008near,silvestri2015online} are not suitable for our case with resource utilization data. This justifies the need of developing our own clustering mechanism that focuses more on short-term spatial correlations. 

More detailed comparison between our approach and the Gaussian-based approach in~\cite{silvestri2015online} will also be presented later in Section~\ref{exp:gaussian}.

\section{Definitions and System Overview}\label{section:problem-formulation}

We consider a distributed system with $N$ local nodes (machines) generating resource utilization measurements, and a central node (controller) that receives a summary of all the local measurements and forecasts the future.
We assume that time is slotted. For each time step $t$, let $\mathbf{x}_t :=[x_{1,t},x_{2,t},...,x_{N,t}]$ denote the $N$-tuple that contains the true measurements of $N$ local nodes and let $\mathbf{z}_t :=[z_{1,t},z_{2,t},...,z_{N,t}]$ be the measurements stored at the central node. 
Here, $x_{i,t}$ and $z_{i,t}$ ($1\leq i \leq N$) are $d$-dimensional vectors, where $d$ is equal to the number of resource types (e.g., CPU, memory).
The values in $\mathbf{z}_t$ depend on the transmission frequency (i.e., how often each local node sends its measurement to the central node). For each node $i$, let $\beta_{i,t}$ be an indication variable such that $\beta_{i,t} = 1$ if node $i$ has sent its most recent measurement at time step $t$ to the central node, otherwise $\beta_{i,t} = 0$. Then, $z_{i,t}=x_{i,t-p}$, where $p\geq 0$ is defined as the smallest $p$ such that $\beta_{i,t-p} = 1$. If $\beta_{i,t} = 1$, then $p=0$ and $z_{i,t}=x_{i,t}$.

We define $K$ as a given input parameter to the system that specifies the number of different forecasting models the system uses, which is related to the computational overhead.
At each time step $t$, the central node partitions the $N$ measurements $z_{1,t},z_{2,t},...,z_{N,t}$ into $K$ clusters, so that one forecasting model can be used for each cluster. Let $C_{j,t}$ ($1\leq j \leq K$) denote the $j$-th cluster at time step $t$, which is defined as a set of indices of local nodes whose measurements are included in this cluster, i.e., $C_{j,t} \subseteq \{1,2,...,N\}$. 
Each cluster $j$ has a centroid, defined as 
\begin{equation}
c_{j,t} := \frac{1}{\left| C_{j,t}  \right|} \sum_{i \in C_{j,t}} z_{i,t}    
\end{equation}
where $|\cdot |$ denotes the cardinality (size) of the set.

\begin{figure}
  \centering
    \includegraphics[width=1\columnwidth]{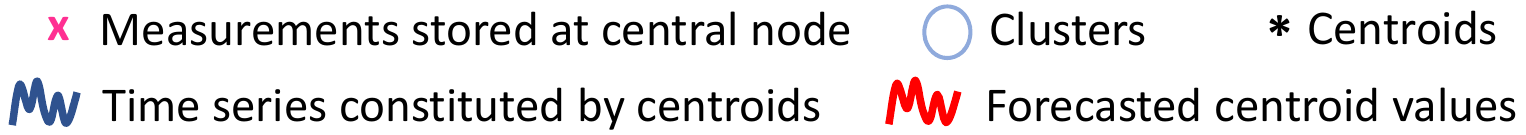}\vspace{0.05in} \\
    \includegraphics[width=1\columnwidth]{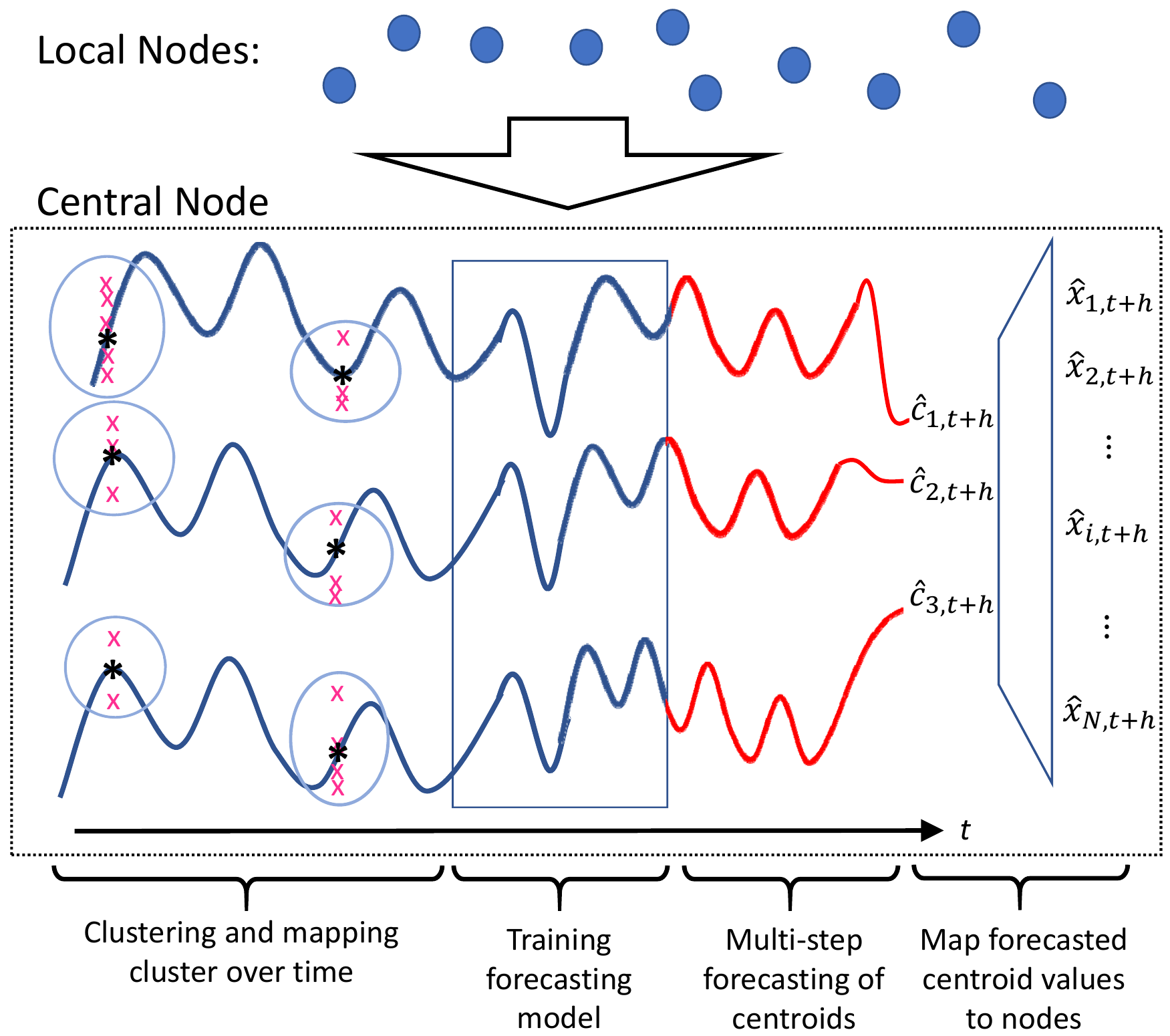}
    \caption{System overview.}
    \label{fig:system-overview}
\end{figure}

At time step $t$, a time-series forecasting model is trained using the time series formed by the set of historical centroids (i.e., $\{c_{j,\tau} : \tau \leq t\}$) for each cluster $j$.
The model can forecast future values of the cluster centroid, i.e., for any forecasting step $h \geq 1$, the model provides a forecasted value $\hat c_{j,{t+h}}$ at the future time step $t+h$. The future resource utilization at each individual local node $i$ is predicted as the value of its centroid plus an offset for this node, thus we define\footnote{For convenience (and with slight abuse of notation), we use the subscript $t+h$ to denote that the current time step is $t$ and we forecast $h$ steps ahead. With this notation, even if $t_1+h_1 = t_2+h_2$, we may have $\hat x_{i,{t_1+h_1}}\neq \hat x_{i,{t_2+h_2}}$ if $t_1 \neq t_2$.}
\begin{align}
\hat x_{i,{t+h}}= \hat c_{j,{t+h}} + \hat s_{i,t+h}
\end{align}
for $i \in \hat C_{j,t+h}$, where $\hat C_{j,t+h}$ is the forecasted set of nodes in cluster $j$ at time step $t+h$, and $\hat s_{i,t+h}$ is the forecasted offset of node $i$ with respect to the centroid of cluster $j$ (to which node $i$ is forecasted to belong to) at time step $t+h$. 
In this way, the estimation of $\hat x_{i,{t+h}}$ involves both spatial estimation\footnote{The use of the term \emph{spatial} estimation or \emph{spatial} correlation is for notional convenience. We acknowledge that the clustering behavior of the measurement data from different local nodes result from their spatial relationship as well as non-spatial reasons such as application-driven workloads.} (using cluster centroid and per-node offset as estimation of values for individual nodes) and temporal forecasting.
Fig.~\ref{fig:system-overview} illustrates the system with the functionalities described above.

We define the root mean square error (RMSE) of $\hat{\mathbf{x}}_{t+h} :=[\hat x_{1,t+h},\hat x_{2,t+h},...,\hat x_{N,t+h}]$ for $h \geq 0$ as
\begin{equation}
\mathrm{RMSE}(t, h) :=\sqrt{ \frac{1}{N} \sum_{i=1}^{N}\left\Vert\hat x_{i,{t+h}}-{x_{i,{t+h}}}\right\Vert^2}
\label{eq:RMSEDef}
\end{equation}
where we define $\hat x_{i,t} := z_{i,t}$ for $h=0$ for convenience. 
With this definition, when $h=0$, the RMSE only includes the error caused by infrequent transmission of local node measurements to the central node.
We also note that the true value $\mathbf{x}_{t+h}$ cannot be observed by the central node.

We also define the time-averaged RMSE over $T$ time steps for a given forecasting step $h$ as
\begin{equation}
\overline{\mathrm{RMSE}}(T, h) :=\sqrt{\frac{1}{T} \sum_{t=1}^T  (\mathrm{RMSE}(t, h))^2}
\label{eq:RMSETimeAverageDef}
\end{equation}
where the time average is over the square error and the square root is taken afterwards.

Let $B_i$ ($0 \leq B_i \leq 1$) denote the maximum transmission frequency (for node $i$).
Using the above definitions, and considering a maximum forecasting range $H$, the algorithms to be introduced in the next section aim at solving the following problem:
\begin{align}
\min & \quad \lim_{T\rightarrow \infty} \sqrt{\frac{1}{H+1} \sum_{h=0}^H (\overline{\mathrm{RMSE}}(T, h))^2} \label{eq:optimizationProblem} \\
\text{s.t.} & \quad \lim_{T\rightarrow \infty} \frac{1}{T} \sum_{t=1}^T \beta_{i,t} \leq B_i, \quad \forall i  \nonumber
\end{align}
where the minimization is over all $\{ \beta_{i,t}\}$, $\{C_{j,t} \}$, $\{\hat C_{j,t+h}\}$,  $\{\hat c_{j,{t+h}}\}$, and $\{\hat s_{i,t+h}\}$.
Intuitively, we would like to find the transmission schedule (indicator) $\beta_{i,t}$ for each local node $i$ and time step $t$, the membership of clusters $C_{j,t}, \forall j$ for each time step $t$, and the forecasted cluster memberships, centroids, and offsets for every forecasting step $h\in [0,H]$ computed at each time step $t$, to minimize the average RMSE over all forecasting steps and all time steps. 

As we do not make any assumption on the characteristics of the time series constituting the cluster centroids $\{c_{j,t}\}$, we cannot hope to find the theoretically optimal forecasting scheme, because for any forecasted time series, there can always exist a true time series that is very different from the forecasted values and thus gives a high forecasting error. In addition, it is often reasonable in the clustering step to minimize the error between the data and their closest cluster centroids (we refer to this error as the ``intermediate RMSE'' later in the paper), which is the K-means clustering problem and is NP-hard~\cite{Aloise2009}.
We also note that an online algorithm is required because measurements from local nodes are obtained over time and decisions have to be made only based on the current and past information (with future information unknown to the algorithm). 
All the above impose challenges in solving (\ref{eq:optimizationProblem}). 
We propose online heuristics to solve the problem (\ref{eq:optimizationProblem}) approximately in the next section. These heuristics work well in practice as we show in Section~\ref{section: experiments} later.

\section{Proposed Algorithms}\label{section:proposed-framework}

\subsection{Measurement Collection with Adaptive Transmission}\label{section:data-collection}

In every time step $t$, each node $i$ determines its action $\beta_{i,t}$, i.e., whether it transmits its current measurement $x_{i,t}$ to the central node or not. 
To capture the error of the measurements stored at the central node, we define a \emph{penalty function}
\begin{equation}
F_{i,t}\left(\beta_{i,t}\right) := 
\begin{cases}
\frac{1}{d} \left\Vert z_{i,t} - x_{i,t}  \right\Vert^2, & \textrm{if }\beta_{i,t} = 0\\
0, & \textrm{if }\beta_{i,t} = 1\\
\end{cases}.
\label{eq:penaltyFunction}
\end{equation}
To take into account the maximum transmission frequency $B_i$, we also define $Y_i \left(\beta_{i,t}\right) := \beta_{i,t} - B_i$.
We also define $V_0>0$ and $\gamma\in (0,1)$ as a control parameters.
The algorithm that runs at each node $i$ to determine $\beta_{i,t}$ is given as follows.
\begin{enumerate}
    \item In the first time slot $t=1$, initialize a variable $Q_i(t) \leftarrow 0$. The variable $Q_i(t)$ represents the length of a ``virtual queue'' at node $i$.
    \item For every $t \in \{ 1, 2, 3,... \}$, choose $\beta_{i,t}$ according to
    \begin{equation}
    \beta_{i,t} \leftarrow \arg\min_{\beta\in\{0,1\}} V_{t} F_{i,t}(\beta) + Q_i(t)Y_i(\beta)
     \label{eq:chooseBeta}
    \end{equation}
    where
    \begin{equation}
    V_{t}:={V_{0}\cdot (t+1)}^\gamma.
    \label{eq:VtDef}
    \end{equation}
    Then, update the virtual queue length according to
    \begin{equation}
    Q_i(t+1) \leftarrow  Q_i(t) + Y_i(\beta_{i,t}). \label{eq:virtualQueueUpdt}
    \end{equation}
\end{enumerate}

The intuition behind the above algorithm is as follows. The virtual queue length $Q_i(t)$ captures how much the $B_i$ constraint in (\ref{eq:optimizationProblem}) has been violated up to the current time step $t$. The determination of $\beta_{i,t}$ in (\ref{eq:chooseBeta}) considers a trade-off between the penalty (error) $F_{i,t}(\beta)$ and constraint violation (related to $Q_i(t)$), where the trade-off is controlled by the parameter $V_t$. When $Q_i(t)$ is large, the term $Q_i(t)Y_i(\beta)$ in (\ref{eq:chooseBeta}) becomes dominant, and the algorithm tends to choose $\beta = 0$ because this gives a negative value of $Q_i(t)Y_i(\beta)$ which is in favor of the minimization. Since $\beta = 0$ corresponds to not transmitting, this relieves the constraint violation. When $Q_i(t)$ is small and $\left\Vert z_{i,t} - x_{i,t}  \right\Vert^2$ is relatively large, the term $V_t F_{i,t}(\beta)$ in (\ref{eq:chooseBeta}) is dominant. In this case, the algorithm tends to choose $\beta = 1$ because this will make $F_{i,t}(\beta) = 0$ and reduces the error of measurements stored at the central node.

The above algorithm is a form of the drift-plus-penalty framework in Lyapunov optimization~\cite{neely2010stochastic}. According to Lyapunov optimization theory, as long as $F_{i,t}(\beta)$ has a finite upper bound\footnote{$F_{i,t}(\beta)$ usually has a finite upper bound because measurement data is usually finite. Also note that the lower bound of $F_{i,t}(\beta)$ is zero thus finite.}, the above algorithm can always guarantee that the $B_i$ constraint in (\ref{eq:optimizationProblem}) is satisfied with equality (for $T \rightarrow \infty$ as given in the constraint, not necessarily for finite $T$), because $\lim_{t \rightarrow \infty} Q_i(t) / t = 0$ (see \cite[Chapter 4]{neely2010stochastic}). Note that satisfying the $B_i$ constraint with equality is always not worse than satisfying it with inequality, because more transmissions cannot hurt the RMSE performance.
For finite $T$, the satisfaction of the $B_i$ constraint is related to the parameter $V_t$, which can be tuned by parameters $V_0$ and $\gamma$. From (\ref{eq:VtDef}), we see that $V_t$ increases with $t$, which means that we give more emphasis on minimizing the penalty function when $t$ is large. This is because for a larger $t$, we can allow a larger $Q_i(t)$ while still maintaining $Q_i(t) / t$ close to zero.

Note, however, that the penalty function $F_{i,t}(\beta)$ depends on transmission decisions in previous time steps that impact the value of $z_{i,t}$. Therefore, the optimality analysis of Lyapunov optimization theory does not hold for our algorithm, and we do not have a theoretical bound on how optimal the result is.
Nevertheless, we have observed that this algorithm with the current penalty definition works well in practice (see experimentation results in Section~\ref{section: experiments}).

\subsection{Dynamic Cluster Construction Over Time}\label{section:dynamic-cluster}

We now discuss how the central node computes the clusters $C_{j,t}$, for $1 \leq j \leq K$, from $\mathbf{z}_t$ over time. 
The computation includes two steps. First, K-means clustering is computed using the stored measurements $\mathbf{z}_t$ in time step $t$ only. Second, the clusters computed in the first step are re-indexed so that they align the best with the clusters computed in previous time steps. The re-indexing step is only performed for $t>1$.

The first step of K-means clustering is straightforward and efficient heuristic algorithms for K-means exist~\cite{KMeansAlg}. Let $C'_{k,t}$ ($1 \leq k \leq K$) denote the K-means clustering result on $\mathbf{z}_t$ in time step $t$. If $t=1$, we let $j=k$, such that $C_{j,t} = C'_{j,t}, \forall j$, where we recall that $\{C_{j,t}: \forall j\}$ is the final set of clusters in time step $t$. If $t>1$, the cluster indices of $\{C'_{k,t}: \forall k\}$ need to be reassigned in order to obtain $\{C_{j,t}: \forall j\}$, because the cluster indices resulting from the K-means algorithm is random, and for each cluster $C'_{k,t}$, we need to find out which cluster among $\{C_{j,t-1}: \forall j\}$ in the previous time step $t-1$ it evolves from.

To associate the clusters $\{C'_{k,t}: \forall k\}$ in time step $t$ with the clusters in previous time steps, we define a similarity measure between the $k$-th cluster from the K-means result in time step $t$, i.e., $C'_{k,t}$, and the $j$-th clusters in a subset of previous time steps. Formally, the similarity measure is defined as
\begin{equation}
w_{k,j}=\left|C'_{k,t} \cap \left(\bigcap_{m=1}^{\min\{M, t-1\}}C_{j,t-m}\right)\right|
\label{eq:similarityMeasure}
\end{equation}
where $M\geq 1$ specifies the number of time steps to look back into the history when computing the intersection in the similarity measure. Intuitively, the similarity measure $w_{k,j}$ specifies how many local nodes exist concurrently in the $k$-th cluster obtained from the K-means algorithm in time step $t$ and in the $j$-th clusters in all $M$ most recent time steps (excluding time step $t$). If $w_{k,j}$ is large, it means that most of the nodes in the corresponding clusters are the same.

Now, to find $C_{j,t}$ from $C'_{k,t}$, we find a one-to-one mapping between the indices $j$ and $k$. Let $\varphi$ denote the one-to-one mapping from $k$ to $j$. We would like to find the mapping $\varphi$ such that the sum similarity is maximized, i.e.,
\begin{equation}
\max_{\varphi} \sum_{k=1}^K w_{k, \varphi(k)}.
\label{eq:clusterMappingProblem}
\end{equation}
Intuitively, with the mapping $\varphi$ found from (\ref{eq:clusterMappingProblem}), the clusters $\{C_{j,t}: \forall j\}$ are indexed in such a way that most nodes remain in the same cluster in the current time step $t$ and $M$ previous time steps. In this way, the evolution of the centroids of each cluster $j$ represents a majority of local nodes within that cluster, and it is reasonable to perform time-series forecasting with the centroids of clusters that are dynamically constructed in this way.

\textbf{Solution to (\ref{eq:clusterMappingProblem}):}
The problem in (\ref{eq:clusterMappingProblem}) is equivalent to a maximum weighted bipartite graph matching problem, where one side of the bipartite graph has nodes representing the values of $k$, the other side of the bipartite graph has nodes representing the values of $j$, and each $k$-$j$ pair is connected with an edge with weight $w_{k,j}$. This can then be solved in polynomial time using existing algorithms for maximum weighted bipartite graph matching, such as the Hungarian algorithm~\cite{kuhn1955hungarian}.

The parameter $M$ in the similarity measure (\ref{eq:similarityMeasure}) controls whether to consider long or short term history when computing the similarity. The proper choice of $M$ is related to the temporal variation in the data correlation among different local nodes, because each cluster contains a group of nodes that are (positively) correlated with each other. Our experimentation results in Section~\ref{section: experiments} show that a fixed value of $M$ usually works well for a given scenario.

Our clustering approach can be extended in several ways. For example, one can define a time window of a given length, which contains multiple time steps, and perform clustering on extended feature vectors that include measurements at multiple time steps within each time window~\cite{WARRENLIAO20051857}. In this case, $t$ represents the time window index, and everything else in our approach presented above works in the same way. We mainly focus on dynamic settings where the time series and node correlation can fluctuate frequently in this paper. In such settings, as we will see in the experimentation results in Section~\ref{section: experiments}, it is best to use a time window of length one (equivalent to no windowing), so that the clustering can adapt to the most recent measurements. 
We can also perform clustering on each type of resource (e.g., CPU, memory) independently from other resource types, in which case the K-means step is performed on one-dimensional vectors (equivalent to scalars). We will see in Section~\ref{section: experiments} that this way of independent clustering performs better than joint clustering on the datasets we use for evaluation.

Our dynamic clustering approach shares some similarities with the approach in~\cite{greene2010tracking}.
However, we define a different similarity measure that can look back multiple time steps and is not normalized. This is more suitable for the RMSE objective in (\ref{eq:optimizationProblem}) which considers the errors at all nodes. Moreover, we focus on the clustering and forecasting of time-series data which is different from existing work.

\subsection{Temporal Forecasting} \label{sec:temporalForecasting}

As discussed in Section~\ref{section:problem-formulation}, temporal forecasting is performed using models trained on historical centroids of measurements stored at the central controller. The models can include Autoregressive Integrated Moving Average (ARIMA) \cite{box1974some}, Long Short-Term Memory (LSTM) \cite{hochreiter1997long}, etc. 
Different models have different computational complexities.
When the system starts for the first time, there is an initial data collection phase where there is no forecasting model available to use. Afterwards, forecasting models are trained on the time-series constituted by the historical centroids of clusters. After the models are trained, the system can forecast future centroids using the models, based on the most updated measurements at the central node. The transient state of each model gets updated whenever a new measurement is available. The models are retrained periodically at a given time interval using all (or a subset of) the historical cluster centroids up to the current time. 

As explained in Section~\ref{section:problem-formulation}, at time step $t$, the forecasted resource utilization at node $i$ in the future time step $t+h$ is computed using the forecasted centroid plus an offset, i.e., $\hat x_{i,{t+h}} = \hat c_{j,t+h} + \hat s_{i,t+h}$ where $j$ is chosen such that $i \in \hat C_{j,t+h}$.
We explain how to find the forecasted cluster $\hat C_{j,t+h}$ and the offset $\hat s_{i,t+h}$ in the following. We define $M'$ as the number of time steps to look back into the history (excluding the current time step $t$). For each node $i$, consider the time steps within the interval $[t-M', t]$, and compute the frequency that node $i$ belongs to the $j$-th cluster $C_{j,t}$ within this time interval, for all $j$. Let $j^*$ denote the cluster that node $i$ belongs to \emph{for the most time} within $[t-M', t]$. The algorithm then predicts that node $i$ belongs to the $j^*$-th cluster in time step $t+h$. By finding $j^*$ for all $i$, the forecasted cluster $\hat C_{j,t+h}$ is obtained for all $j$. 

For node $i \in \hat C_{j,t+h}$, the offset $\hat s_{i,t+h}$ is computed as
\begin{equation}
    \hat s_{i,t+h}=\frac{1}{M'+1}\sum_{m=0}^{M'} \alpha_{t-m}(z_{i,t-m}-c_{j,t-m})
\end{equation}
where $\alpha_{t-m} \in (0, 1]$ is a scaling coefficient that ensures the cluster centroid plus the offset $c_{j,t-m} + \alpha_{t-m}(z_{i,t-m}-c_{j,t-m})$ still belongs to cluster $j$ in time step $t-m$, i.e., its value is still closest to the centroid $c_{j,t-m}$ of cluster $j$ compared to the centroids of all other clusters.
If $z_{i,t-m}$ belongs cluster $j$, we choose $\alpha_{t-m}=1$. Otherwise, we choose $\alpha_{t-m}$ as the largest value so that $c_{j,t-m} + \alpha_{t-m}(z_{i,t-m}-c_{j,t-m})$ belongs to cluster $j$.
This is useful because we do not want the offset to be so large that the resulting estimated value belongs to a different cluster (other than cluster $j$), as the forecasted $\hat x_{i,{t+h}}$ is still based on the forecasted centroid $\hat c_{j,t+h}$ of cluster $j$.

\section{Experimentation Results}\label{section: experiments}

\begin{subsection}{Setup}\label{sec:experiment-setup}

We evaluate the performance of our proposed approach on three real-world computing cluster datasets.

\subsubsection{Datasets}

The first dataset is the \emph{Alibaba cluster trace (version 2018)}~\cite{alibaba} that includes CPU and memory utilizations of $4,000$ machines over a period of $8$ days. The raw measurements are sampled at $1$~minute intervals (i.e., each local node obtains a new measurement every minute) and the entire compressed dataset is about $48$~GB.
The second dataset is the Rnd trace of the \emph{GWA-T-12 Bitbrains dataset}~\cite{shen2014workload}. It contains $500$ machines, the data is collected over a period of $3$ months (we only use data in the first month because there is a $24$-hour gap between different months), and raw measurements are sampled at $5$~minute intervals. The size of the dataset is $156$~MB.
The third dataset is the \emph{Google cluster usage trace (version 2)}~\cite{reiss2011google}, which contains job/task usage information of approximately $12,478$~machines\footnote{We had to remove $2$ machines that have error in the measurement data (unreasonably high CPU/memory utilization).} over $29$~days, sampled at $5$~minute intervals. The total size of the compressed dataset is approximately $41$~GB. 
For each dataset, we pre-processed the raw data to obtain the \emph{normalized} CPU and memory utilizations for each individual machine.

\subsubsection{Choice of Parameters}
\label{subsec:experiment-parameters}
Unless otherwise specified, we set the transmission frequency constraint $B_i = B := 0.3$ 
for all $i$, the control parameters for adaptive transmission  $V_{0} ={10^{-12}}$ and $\gamma=0.65$, the number of forecasting models (which is equal to the number of clusters) $K=3$, the look-back durations for the similarity measure  $M=1$  and temporal forecasting $M'=5$. The clustering is performed on the scalar values of the measurements of each resource type, unless noted otherwise. 
These parameter choices are justified in our experiments, which will be further discussed later in this section.

\begin{figure}[t]
    \centering
       \begin{subfigure}[b]{0.28\columnwidth}
        \centering
        \includegraphics[width=1\linewidth]{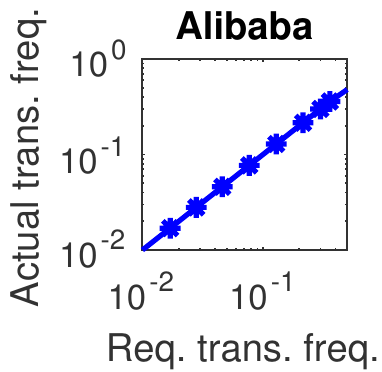}
       
    \end{subfigure}%
    ~~~
    \begin{subfigure}[b]{0.28\columnwidth}
        \centering
        \includegraphics[width=1\linewidth]{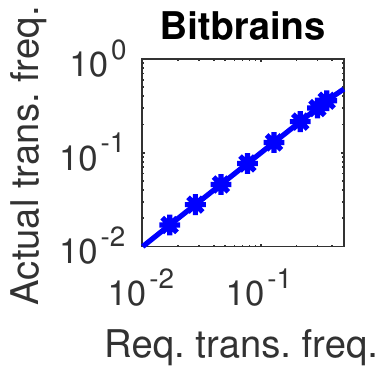}
      
    \end{subfigure}%
    ~~~
    \begin{subfigure}[b]{0.28\columnwidth}
        \centering
        \includegraphics[width=1\linewidth]{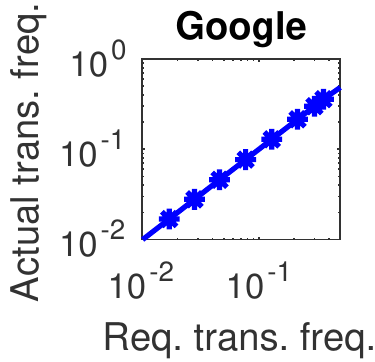}
     
    \end{subfigure}%

    \caption{Behavior of the adaptive transmission algorithm}
    \label{figure:control}
\end{figure}

\begin{figure*}[t]
    \centering

     \begin{subfigure}{0.45\columnwidth}
        \centering
        \includegraphics[width=1\linewidth]{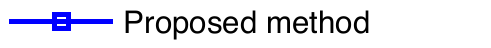}
    \end{subfigure}%
    \quad
     \begin{subfigure}{0.45\columnwidth}
        \centering
        \includegraphics[width=1\linewidth]{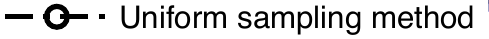}
    \end{subfigure}%
    \vspace{0.05in}

    \begin{subfigure}{0.33\columnwidth}
        \centering
        \includegraphics[width=1\linewidth]{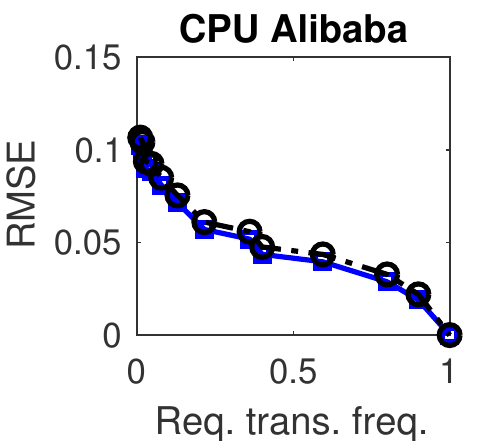}
    \end{subfigure}%
  \begin{subfigure}{0.33\columnwidth}
        \centering
        \includegraphics[width=1\linewidth]{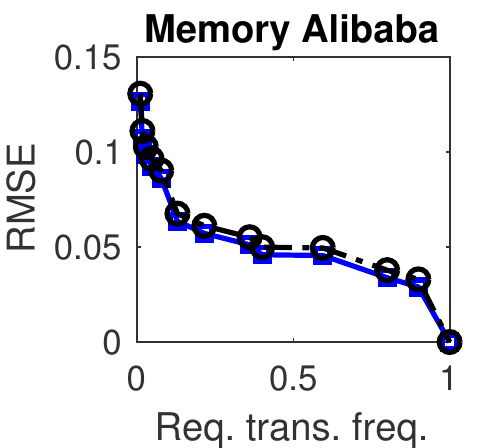}
    \end{subfigure}%
    \begin{subfigure}{0.33\columnwidth}
        \centering
        \includegraphics[width=1\linewidth]{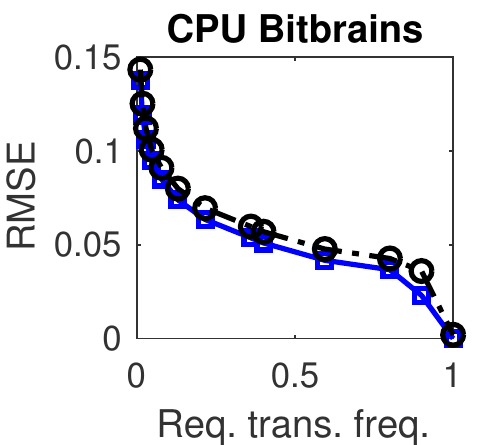}
    \end{subfigure}%
    \begin{subfigure}{0.33\columnwidth}
        \centering
        \includegraphics[width=1\linewidth]{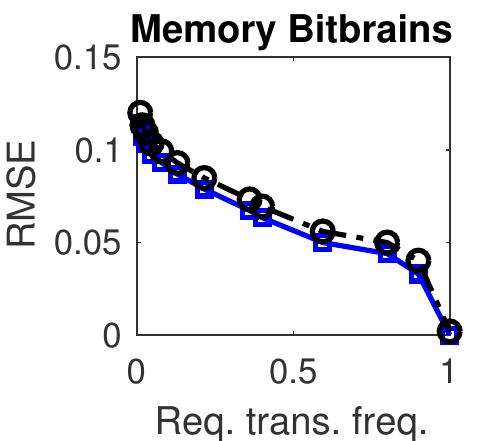}
    \end{subfigure}%
    \begin{subfigure}{0.33\columnwidth}
        \centering
        \includegraphics[width=1\linewidth]{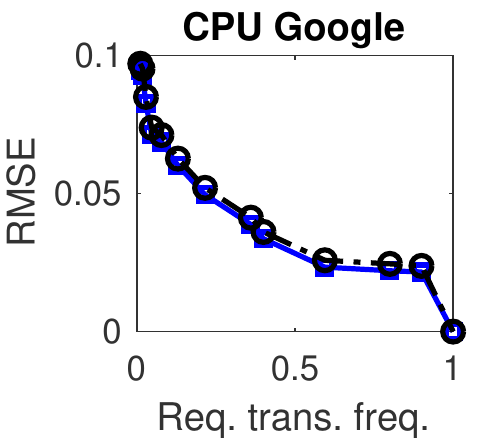}
    \end{subfigure}%
     \begin{subfigure}{0.33\columnwidth}
        \centering
        \includegraphics[width=1\linewidth]{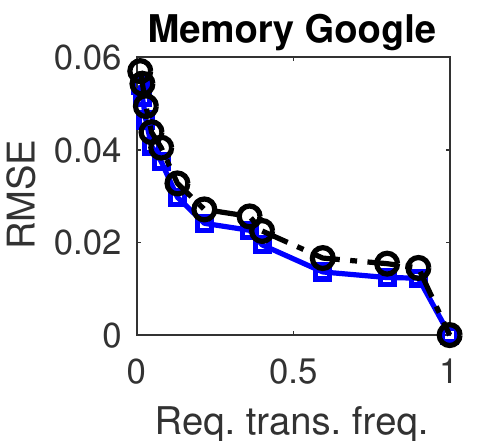}
    \end{subfigure}%
    
    \caption{RMSE comparison of our proposed adaptive transmission method with the uniform sampling method}
    \label{figure:comparison-naive}
    \vspace{-0.15in}
\end{figure*}

\subsubsection{Forecasting Models} \label{subsec:experiment-forecasting-model-definition}
We use ARIMA and LSTM models for temporal forecasting.
For the ARIMA model, after making some initial  observations of the stationarity, auto correlation, and partial auto correlation functions, we conduct a grid search over 
the following ranges of parameters: the order of the auto-regressive terms $p \in [0,5]$, the degree of differencing $d \in [0,2]$, the order of the moving average terms $q \in [0,5]$, and for the corresponding seasonal components: $P \in [0,2]$, $D \in [0,1]$, $Q \in [0,2]$.
The best model is selected from the grid search using the Akaike information criterion with correction term (AICc) \cite{burnham2003model}.
For the LSTM model, we stacked two LSTM layers, and on top of that we stacked a dense layer with a rectified linear unit (ReLU) as activation function. Due to the randomness of LSTM, we plot the average forecasting results over $10$ different simulation runs.

For both ARIMA and LSTM, the initial data collection phase includes the first $1000$ time steps. Then, the models are retrained every $288$ time steps, equivalent to a day when the raw measurements are sampled at $5$~minute intervals. For each cluster $j$, a separated model is trained for forecasting the centroids of this cluster.
At every time step $t$, forecasting is made for a given number of time steps $h$ ahead.

We present results on different aspects of our proposed mechanism in the following.

\emph{Remark:} As mentioned in Section~\ref{section:related}, to the best of our knowledge, there does not exist work in the literature that solves the entire problem in our setting. Therefore, we cannot compare our overall method with another existing approach. We will compare individual parts of our method with existing work where possible.

\end{subsection}

\subsection{Adaptive Transmission Algorithm}\label{transmission}
We first study some behavior of the algorithm presented in Section~\ref{section:data-collection}. Fig.~\ref{figure:control} shows that the required transmission frequency $B$ always matches closely with the actual transmission frequency (with parameters $V_0$ and $\gamma$ chosen as described in Section~\ref{subsec:experiment-parameters}).  This confirms that the algorithm is able to adapt the transmission frequency to remain within the $B_i$-constraint in (\ref{eq:optimizationProblem}).

In Fig.~\ref{figure:comparison-naive}, we compare our proposed adaptive transmission approach with a uniform sampling approach, and show the time-averaged RMSE as defined in (\ref{eq:RMSETimeAverageDef}) with $h=0$ and $T$ equal to the total number of time steps in the dataset (recall that we defined $\hat x_{i,t} := z_{i,t}$ for $h=0$, so the RMSE only includes error caused by infrequent transmission in this case). The \emph{uniform sampling} baseline transmits each local node's measurement at a fixed interval, so that the average transmission frequency at each node $i$ is equal to $B_i$. We see that our proposed approach outperforms the uniform sampling approach for any required transmission frequency. When the required transmission frequency is $1.0$, we always have $z_{i,t} = x_{i,t}$ and the RMSE is zero for both approaches.

\subsection{Spatial Estimation without Per-node Offset}

In this subsection, we evaluate the impact of using cluster centroids to represent the group of nodes in the cluster, where we \emph{ignore the offset} $\hat s_{i,t+h}$ and choose $h=0$. We evaluate the \emph{intermediate RMSE} which is the time-averaged RMSE between the data and their closest cluster centroids. This evaluation is useful because the forecasting models are trained on cluster centroids, so we would like the cluster centroids to be not too far from the actual measurements at each node even if there is no per-node offset added to the estimated value. It also provides useful insights on the clustering mechanism.

\subsubsection{Impact of Clustering Dimensions}
\begin{figure}[t]
    \centering
    
      \begin{subfigure}{0.3\columnwidth}
    \centering
        \includegraphics[width=0.8\linewidth]{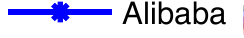}
    \end{subfigure}%
    \quad
     \begin{subfigure}{0.3\columnwidth}
        \centering
        \includegraphics[width=0.8\linewidth]{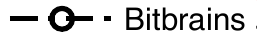}
    \end{subfigure}%
      \quad
     \begin{subfigure}{0.3\columnwidth}
        \centering
        \includegraphics[width=0.8\linewidth]{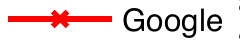}
    \end{subfigure}%

    \begin{subfigure}{0.4\columnwidth}
        \centering
        \includegraphics[width=1\linewidth]{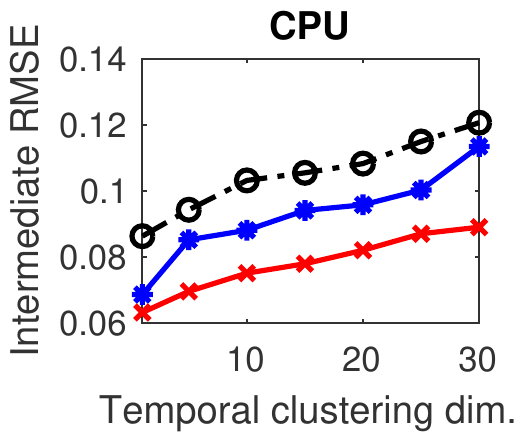}
    
    \end{subfigure}%
    ~~~~
    \begin{subfigure}{0.4\columnwidth}
        \centering
        \includegraphics[width=1\linewidth]{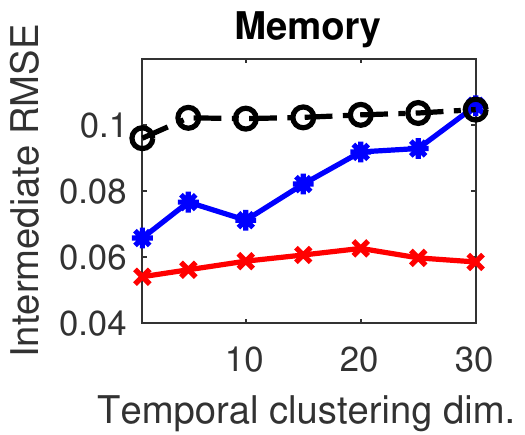}
      
    \end{subfigure}%

    \caption{Intermediate RMSE of clustering different temporal dimensions.}
    \label{figure:varying-temporaldim}
    \vspace{0.1in}
\end{figure}

We first discuss the impact of different dimensions we cluster over time and over resource types.
As mentioned in Section~\ref{section:dynamic-cluster}, we can cluster either on the measurement obtained at a single time step or multiple time steps, i.e., over different temporal dimensions.
Fig.~\ref{figure:varying-temporaldim} shows the results of intermediate RMSE when we vary the temporal clustering dimension, where we cluster CPU and memory measurements separately and independently. We see that using a temporal clustering dimension of $1$ (i.e., clustering the measurements obtained at a single time step) always gives the best performance.

\begin{table}[t]
\caption{Intermediate RMSE of clustering independent scalars \& full vectors \vspace{-0.25in}} {
\footnotesize
\begin{center}
\begin{tabular}{lccc}
\hline
\textbf{Resource type \& dataset} & \textbf{Scalar} & \textbf{Full}\\ \hline
CPU Alibaba          &    0.069                &   0.075    \\ 
Memory Alibaba     &      0.066         &0.072    \\  
CPU Bitbrains          &   0.086                & 0.089        \\ 
Memory Bitbrains     &      0.096            & 0.098 \\ CPU Google          &     0.063                &      0.082  \\ 
Memory Google     &      0.055         & 0.067 \\  \hline
\end{tabular}
\end{center}
}
\label{table:dim-clustering}
\end{table}

\begin{figure*}[t]
   \centering

     \begin{subfigure}[t]{0.33\columnwidth}
        \centering
        \includegraphics[width=1\linewidth]{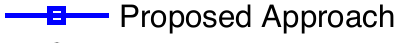}
    \end{subfigure}%
    ~
    \begin{subfigure}[t]{0.33\columnwidth}
        \centering
        \includegraphics[width=1\linewidth]{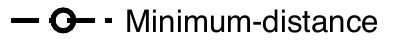}
    \end{subfigure}%
    ~
    \begin{subfigure}[t]{0.33\columnwidth}
        \centering
        \includegraphics[width=1\linewidth]{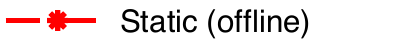}
    \end{subfigure}%
    \vspace{0.05in}

    \begin{subfigure}{0.33\columnwidth}
        \centering
        \includegraphics[width=1\linewidth]{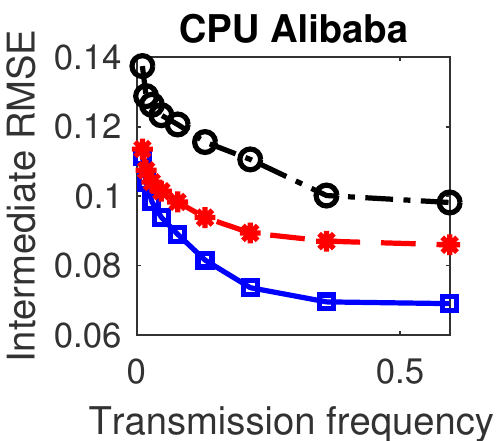}
    \end{subfigure}%
    \begin{subfigure}{0.33\columnwidth}
        \centering
        \includegraphics[width=1\linewidth]{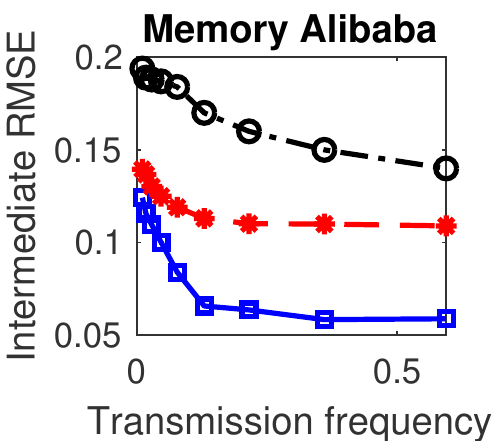}
    \end{subfigure}%
      \begin{subfigure}{0.33\columnwidth}
        \centering
        \includegraphics[width=1\linewidth]{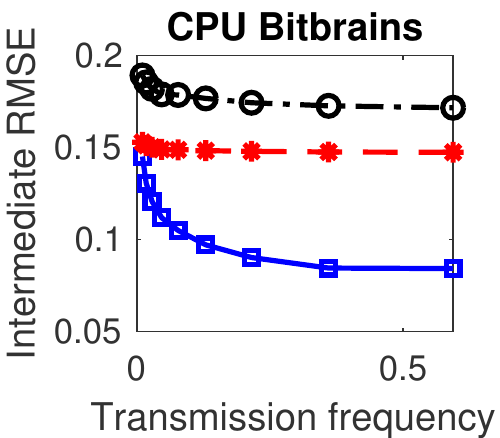}
    \end{subfigure}%
    \begin{subfigure}{0.33\columnwidth}
        \centering
        \includegraphics[width=1\linewidth]{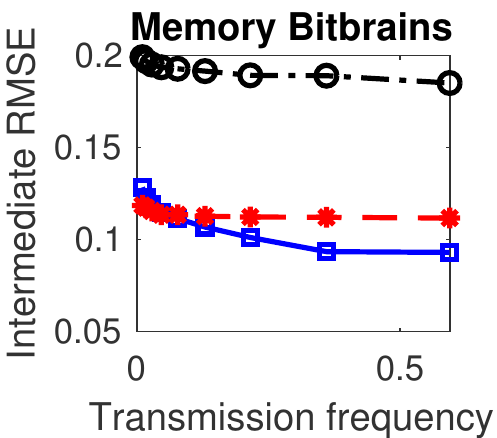}
    \end{subfigure}%
    \begin{subfigure}{0.33\columnwidth}
        \centering
        \includegraphics[width=1\linewidth]{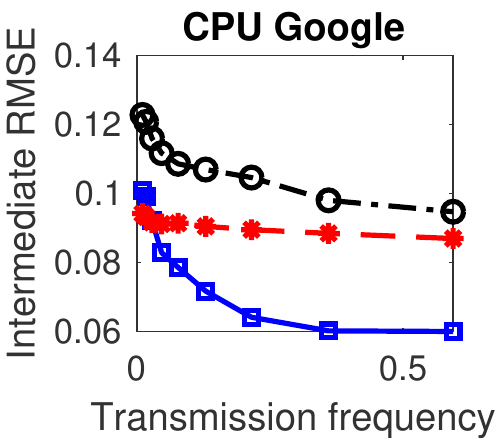}
    \end{subfigure}%
    \begin{subfigure}{0.33\columnwidth}
        \centering
        \includegraphics[width=1\linewidth]{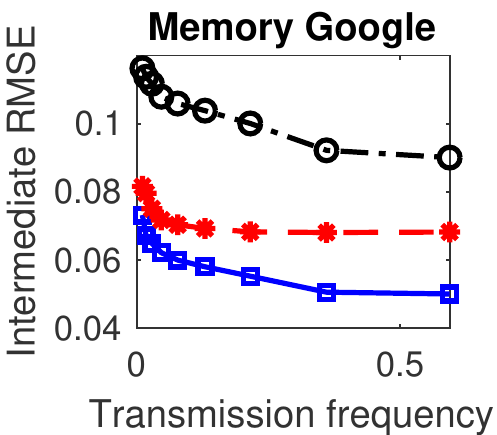}
    \end{subfigure}%

     \caption{Intermediate RMSE when varying the transmission frequency $B$ and fixing $K=3$.} 
    \label{fig:spatialEstVaryTransmissionFreq}
    \vspace{-0.1in}
\end{figure*}

\begin{figure*}[t]
    \centering
    
     \begin{subfigure}[t]{0.33\columnwidth}
        \centering
        \includegraphics[width=1\linewidth]{figure2/legend/legend-propap.pdf}
    \end{subfigure}%
    ~
    \begin{subfigure}[t]{0.33\columnwidth}
        \centering
        \includegraphics[width=1\linewidth]{figure2/legend/legend-min.pdf}
    \end{subfigure}%
    ~
    \begin{subfigure}[t]{0.33\columnwidth}
        \centering
        \includegraphics[width=1\linewidth]{figure2/forecast/legend3-2.pdf}
    \end{subfigure}%
    \vspace{0.05in}

    \begin{subfigure}{0.33\columnwidth}
        \centering
        \includegraphics[width=1\linewidth]{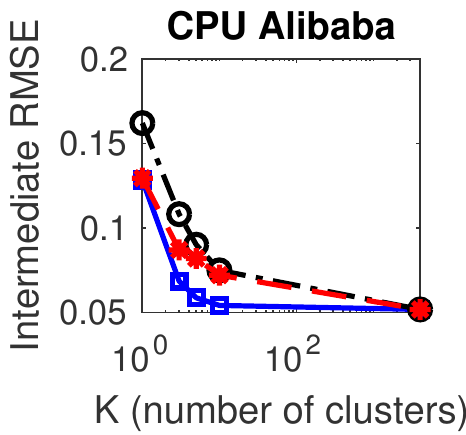}
    \end{subfigure}%
    \begin{subfigure}{0.33\columnwidth}
        \centering
        \includegraphics[width=1\linewidth]{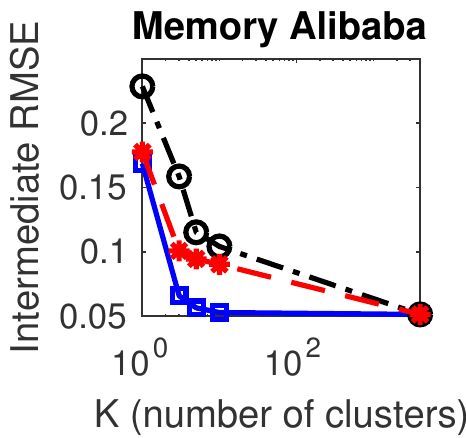}
    \end{subfigure}%
    \begin{subfigure}{0.33\columnwidth}
        \centering
        \includegraphics[width=1\linewidth]{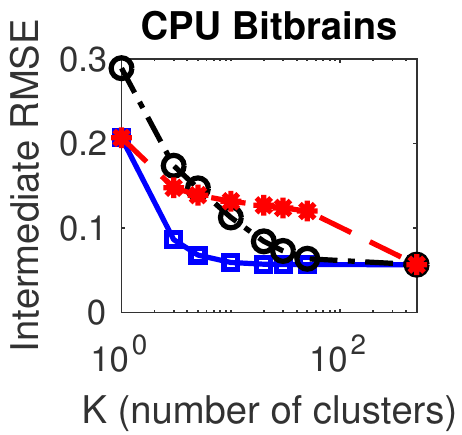}
    \end{subfigure}%
    ~
    \begin{subfigure}{0.33\columnwidth}
        \centering
        \includegraphics[width=1\linewidth]{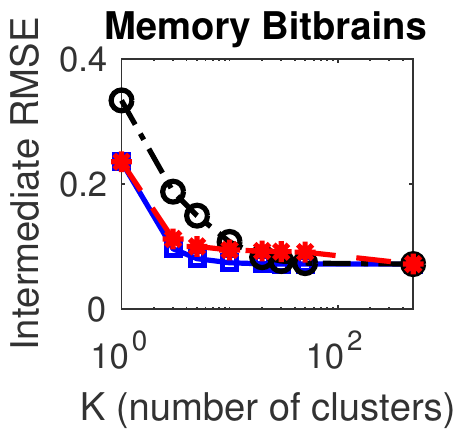}
    \end{subfigure}%
    \begin{subfigure}{0.33\columnwidth}
        \centering
        \includegraphics[width=1\linewidth]{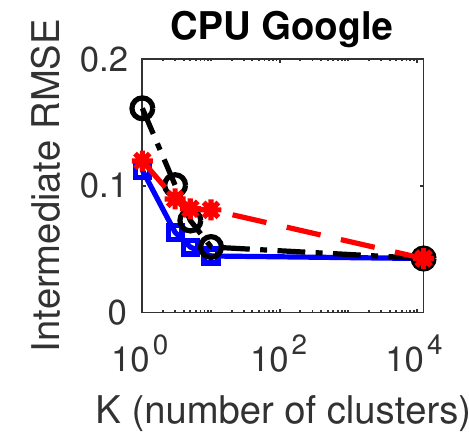}
    \end{subfigure}%
    \begin{subfigure}{0.33\columnwidth}
        \centering
        \includegraphics[width=1\linewidth]{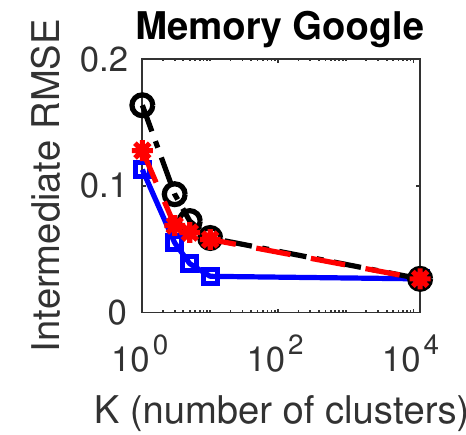}
    \end{subfigure}%
    
    \caption{Intermediate RMSE when varying the number of clusters $K$ and fixing $B=0.3$.} 
    \label{fig:spatialEstVaryK}
    \vspace{-0.1in}
\end{figure*}

\begin{figure*}[t]
    \centering
    \begin{subfigure}[b]{0.4\columnwidth}
        \centering
        \includegraphics[width=0.7\linewidth]{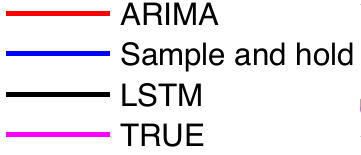}
        \vspace{0.6in}
    \end{subfigure}%
    \begin{subfigure}[b]{0.55\columnwidth}
        \centering
        \includegraphics[width=1\linewidth]{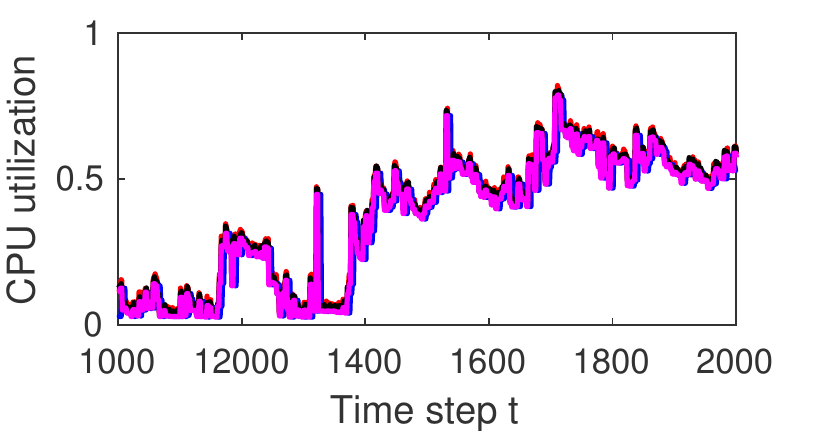}
        \caption{Centroid $j=1$}
    \end{subfigure}%
    \begin{subfigure}[b]{0.55\columnwidth}
        \centering
        \includegraphics[width=1\linewidth]{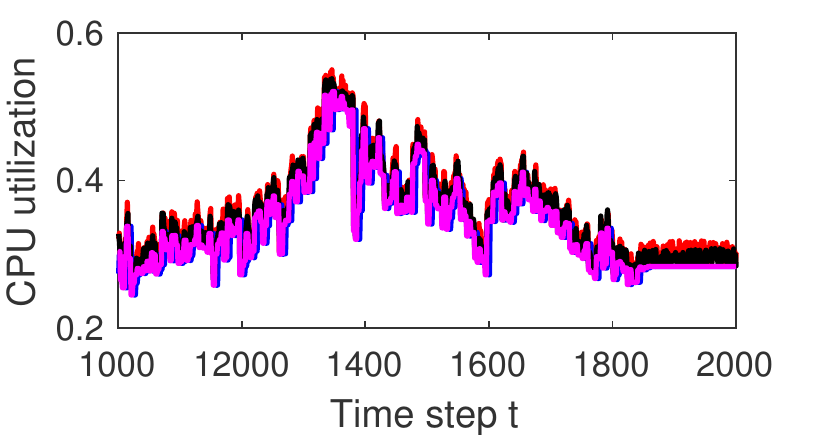}
        \caption{Centroid $j=2$}
    \end{subfigure}%
    \begin{subfigure}[b]{0.55\columnwidth}
        \centering
        \includegraphics[width=1\linewidth]{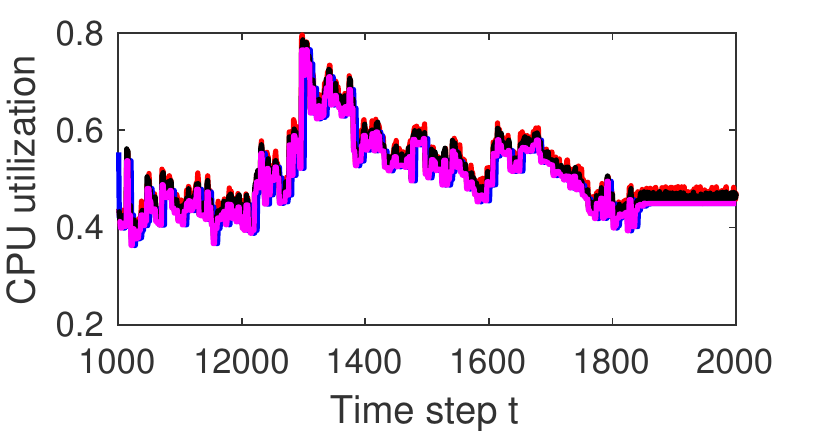}
        \caption{Centroid $j=3$}
    \end{subfigure}%
    
    \caption{Instantaneous true and forecasted ($h=5$) results of $K=3$ centroids on CPU data of Alibaba dataset.}
    \label{fig:ExperimentInstantaneousForecast}
    \vspace{-0.15in}
\end{figure*}

Section~\ref{section:dynamic-cluster} also mentions that we can either cluster different resource types independently using their scalar values, or we can jointly cluster vectors of multiple resource types.
Table~\ref{table:dim-clustering} compares the intermediate RMSEs of these two approaches, where the intermediate RMSEs are always computed for individual resource types, but the clustering is computed either on independent scalars or full vectors. We see that clustering using scalar values of each resource type performs better than clustering using the full vector. This suggests that the correlation among different types of resources in each dataset is relatively weak.

The above results show that it is beneficial to use scalar measurement values of each resource type at a single time step for clustering. We will use this setting in all our experiments presented next.

\subsubsection{Different Clustering Methods} \label{subsubsec:experiment-spatial-estimation-diff-clustering}

We compare our proposed dynamic clustering approach with two baselines. The first baseline \emph{static clustering} is an \emph{offine} baseline, where nodes are grouped into static clusters based on the entire time series at each node that is assumed to be known in advance. The clusters are found using K-means on multi-dimensional vectors, where each vector represents the entire time series at a node. With this setting, the clusters remain fixed over all time steps. The second baseline \emph{minimum distance} is obtained by randomly selecting $K$ nodes at each time step, treating the selected nodes as ``centroids'' and mapping the remaining nodes to the ``centroids'' based on minimum Euclidean distance between measurements. The minimum distance baseline represents approaches which select monitoring nodes randomly, such as~\cite{coluccia2011lossy,barcelo2012enhanced,anagnostopoulos2014advanced,li2018compressed,leinonen2014compressed}.

Fig.~\ref{fig:spatialEstVaryTransmissionFreq} shows the intermediate RMSE with varying $B$ while fixing $K=3$. We can see that our proposed approach performs better than baseline approaches in (almost) all cases. Note that the static approach is an offline baseline with stronger assumptions than our proposed online approach. We also see that in most cases, the intermediate RMSE starts to converge at approximately $B=0.3$. This shows that a transmission frequency higher than $0.3$ will not provide much benefit.

Fig.~\ref{fig:spatialEstVaryK} shows the results with varying $K$ while fixing $B=0.3$. We see that the intermediate RMSE of the proposed approach is close to the lowest value even with only a few clusters (i.e., small value of $K$). This is a strong result because it shows that a small number of cluster centroids is sufficient for representing a large number of nodes.
We also note that because $B=0.3$, the measurements stored at the central node are not always up-to-date, which explains why the intermediate RMSE is larger than zero even when $K=N$.

The above observations explain the rationale behind choosing $B=0.3$ and $K=3$ as default parameters as mentioned in Section~\ref{subsec:experiment-parameters}.
In general, we can conclude that our proposed approach can provide close to optimal clustering error by using a small transmission frequency and a very small number of clusters, which significantly reduces the communication and computation overhead for system monitoring.

\begin{figure*}[t]
    \centering
    
     \begin{subfigure}{0.25\columnwidth}
    \centering
    \includegraphics[width=1\linewidth]{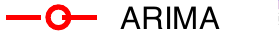}
    \end{subfigure}%
    \quad
    \begin{subfigure}{0.36\columnwidth}
    \centering
    \includegraphics[width=1\linewidth]{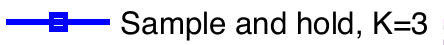}
    \end{subfigure}%
    \quad
     \begin{subfigure}{0.36\columnwidth}
        \centering
        \includegraphics[width=1\linewidth]{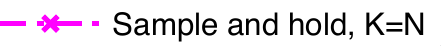}
    \end{subfigure}%
    \quad
     \begin{subfigure}{0.3\columnwidth}
        \centering
        \includegraphics[width=1\linewidth]{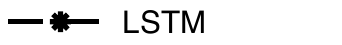}
    \end{subfigure}%
    \quad
     \begin{subfigure}{0.33\columnwidth}
        \centering
        \includegraphics[width=1\linewidth]{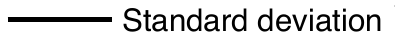}
    \end{subfigure}%

    \vspace{0.05in}
    
    \begin{subfigure}[b]{0.33\columnwidth}
        \centering
        \includegraphics[width=1\linewidth]{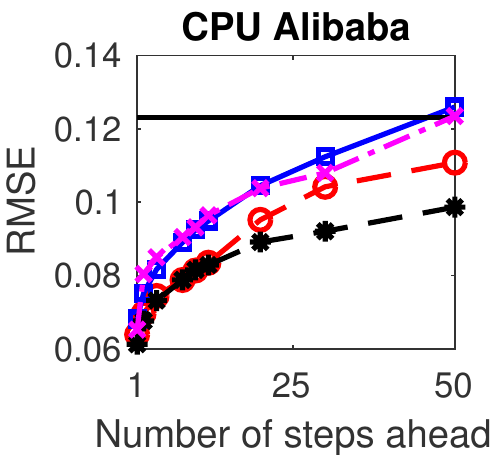}
    \end{subfigure}%
    \begin{subfigure}[b]{0.33\columnwidth}
        \centering
        \includegraphics[width=1\linewidth]{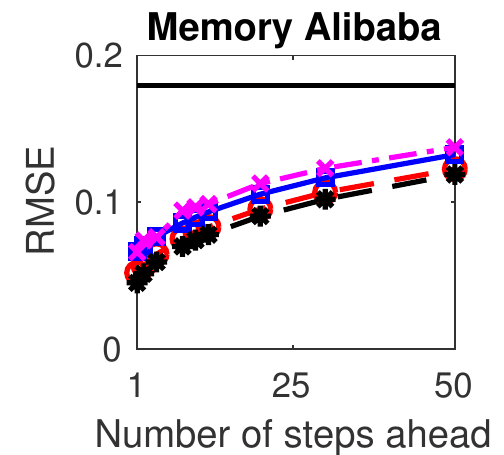}
    \end{subfigure}%
    \begin{subfigure}[b]{0.33\columnwidth}
        \centering
        \includegraphics[width=1\linewidth]{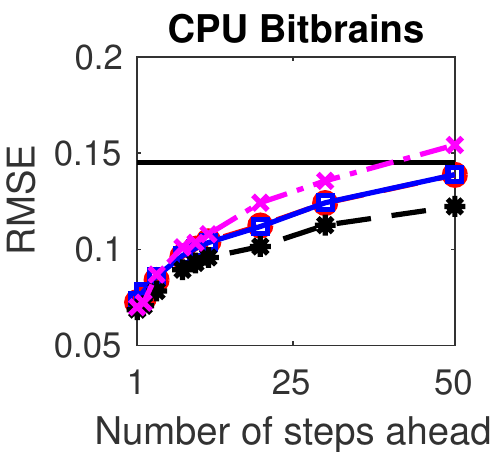}
    \end{subfigure}%
    \begin{subfigure}[b]{0.33\columnwidth}
        \centering
        \includegraphics[width=1\linewidth]{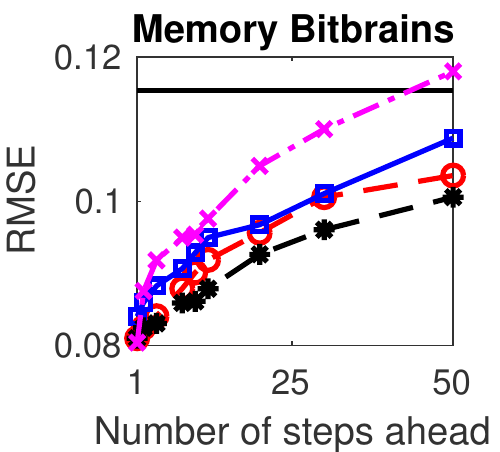}
    \end{subfigure}%
    \begin{subfigure}[b]{0.33\columnwidth}
        \centering
        \includegraphics[width=1\linewidth]{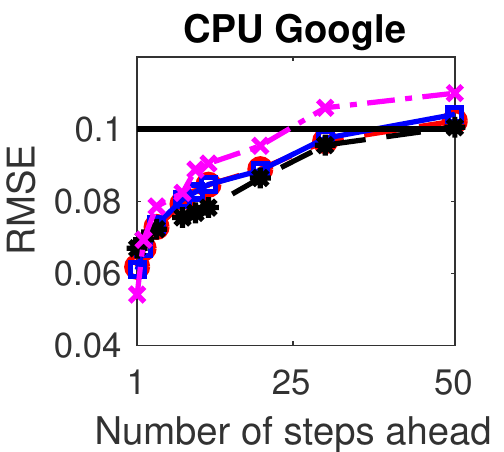}
    \end{subfigure}%
     \begin{subfigure}[b]{0.33\columnwidth}
        \centering
        \includegraphics[width=1\linewidth]{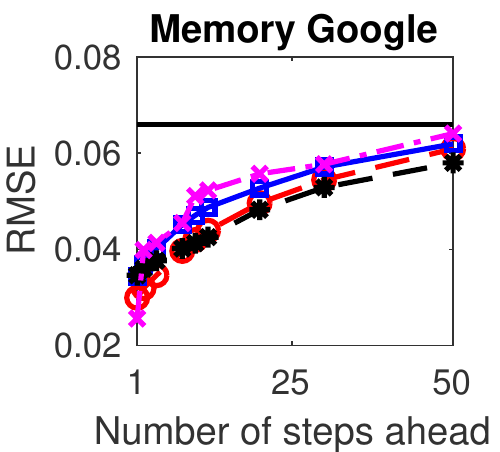}
    \end{subfigure}%
    
    \caption{Time-averaged RMSE with different number of forecasting steps ($h$), with our proposed dynamic clustering approach.}
    \label{figure:forecast2-intel}
    \vspace{-0.1in}
\end{figure*}

\begin{figure*}[t]
    \centering
   \begin{subfigure}{0.35\columnwidth}
    \centering
    \includegraphics[width=1\linewidth]{figure2/legend/legend-propap.pdf}
    \end{subfigure}%
    \quad
    \begin{subfigure}{0.35\columnwidth}
    \centering
    \includegraphics[width=1\linewidth]{figure2/legend/legend-min.pdf}
    \end{subfigure}%
    \quad
     \begin{subfigure}{0.35\columnwidth}
        \centering
        \includegraphics[width=1\linewidth]{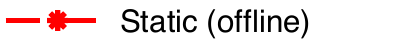}
    \end{subfigure}%
    \quad
     \begin{subfigure}{0.33\columnwidth}
        \centering
        \includegraphics[width=1\linewidth]{figure2/legend/l-std.pdf}
    \end{subfigure}%
    \vspace{0.05in}
    
    \begin{subfigure}[b]{0.33\columnwidth}
        \centering
        \includegraphics[width=1\linewidth]{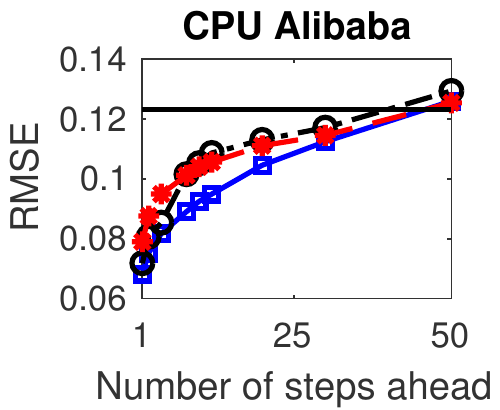}
    \end{subfigure}%
    \begin{subfigure}[b]{0.33\columnwidth}
        \centering
        \includegraphics[width=1\linewidth]{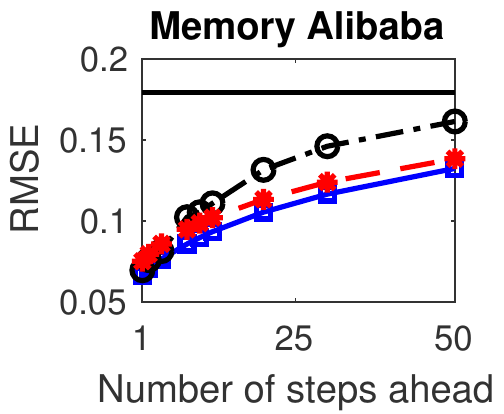}
    \end{subfigure}%
    \begin{subfigure}[b]{0.33\columnwidth}
        \centering
        \includegraphics[width=1\linewidth]{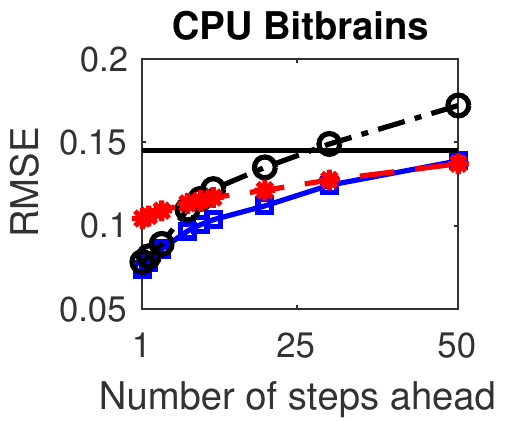}
    \end{subfigure}%
    \begin{subfigure}[b]{0.33\columnwidth}
        \centering
        \includegraphics[width=1\linewidth]{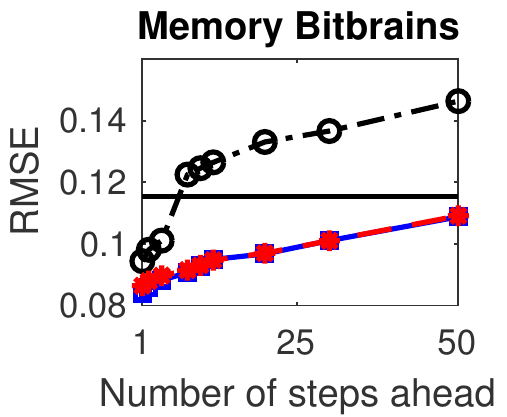}
    \end{subfigure}%
    \begin{subfigure}[b]{0.33\columnwidth}
        \centering
        \includegraphics[width=1\linewidth]{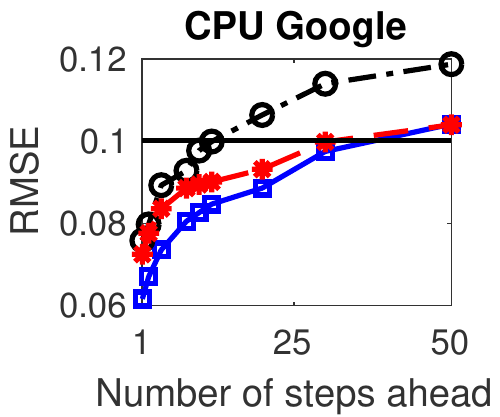}
    \end{subfigure}%
    \begin{subfigure}[b]{0.33\columnwidth}
        \centering
        \includegraphics[width=1\linewidth]{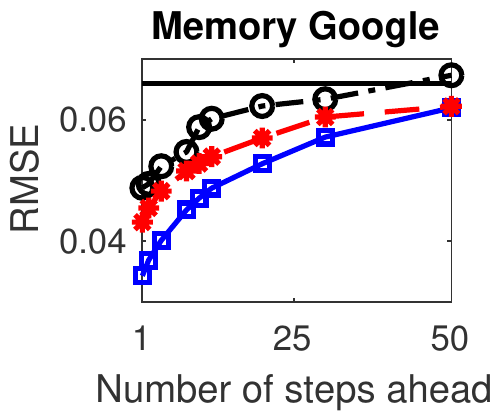}
    \end{subfigure}%
    
    \caption{Time-averaged RMSE with different number of forecasting steps ($h$) using the sample-and-hold method.}
    \label{figure:forecast-sample}
    \vspace{-0.15in}
\end{figure*}

\subsection{Joint Spatial Estimation and Temporal Forecasting (with Per-node Offset)}

We now consider the entire pipeline with joint spatial estimation (through dynamic clustering) and temporal forecasting. We include the per-node offset $\hat s_{i,t+h}$ in this subsection and focus on the time-averaged RMSE as defined in (\ref{eq:RMSETimeAverageDef}).

\subsubsection{Different Forecasting Models}

We compare our predictions based on ARIMA and LSTM with a \emph{sample-and-hold} prediction method, which simply uses the cluster centroid values at time step $t$ as the predicted future values. We also compare with the \emph{standard deviation} computed over all resource utilizations over time (except for the instantaneous plot in Fig.~\ref{fig:ExperimentInstantaneousForecast}). The standard deviation serves as an \emph{error upper bound of an offline mechanism} where forecasting is made only based on long-term statistics (such as mean value) without considering temporal correlation.

We first show the instantaneous true and forecasted CPU utilization values of three different centroids for $t\in[1000,2000]$ with the Alibaba dataset in Fig.~\ref{fig:ExperimentInstantaneousForecast}, where the forecasting is for $h=5$ steps ahead. 
We see that with our methods, the trajectories of the forecasted centroid values by all models follow very closely to that of the true centroid values.

The time-averaged RMSE with different forecasting models is shown in Fig.~\ref{figure:forecast2-intel}, where we include results for both $K=3$ and $K=N$ for the sample-and-hold method, and use the default $K=3$ for all the other methods. Also note that the standard deviation does not depend on $K$.
We see that although sample-and-hold is simple enough to run on every local node (i.e., $K=N$), the case with $K=N$ generally performs worse than cases with $K=3$. This is due to the fluctuation of resource utilization at individual nodes, which makes the forecasting model perform badly when running on every node. The cluster centroids are averages of data at multiple nodes, which remove noisy fluctuations and provide better performances.
LSTM performs the best among all the models, which is expected since LSTM is the most complex and advanced model compared to the others. 
We also see that the RMSE is lower than the standard deviation for most forecasting models when the forecasting step $h\leq 50$. This shows that our forecasting mechanism, which takes into account both spatial and temporal correlations, is beneficial over mechanisms that are only based on long-term statistics.

Table~\ref{tab:TrainingTime} shows the total (aggregated) computation time used for training the ARIMA and LSTM models for the entire duration of one centroid, on a regular personal computer (without GPU) with Intel Core i7-6700 $3.4$~GHz CPU, $16$~GB memory. The model is trained or re-trained at each of the initial training and retraining periods  defined in Section~\ref{subsec:experiment-forecasting-model-definition}, and the result shown in Table~\ref{tab:TrainingTime} is the sum computation time for training at all periods. We can see that for data traces that span over at least multiple days, the total computation time used for model training is only a few minutes. Since we only need to train $K=3$ models, the computation overhead (time) for training forecasting models is very small compared to the entire monitoring duration.

\begin{table}[t]
\caption{Aggregated training time (in seconds) of forecasting model on one centroid over the entire duration of the dataset \vspace{-0.1in}}
\label{tab:TrainingTime} {
\footnotesize
\begin{center}
\begin{tabular}{lll}
\hline
\textbf{Dataset}                                                                     & \textbf{ARIMA} & \textbf{LSTM} \\ \hline
{\begin{tabular}[c]{@{}l@{}}Alibaba data set \\ ($11519$ total time steps)\end{tabular}} & 61.25         & 855.34        \\ \hline
{\begin{tabular}[c]{@{}l@{}}Bitbrains data set\\ ($8259$ total time steps)\end{tabular}} & 33.4          & 554.97        \\ \hline
{\begin{tabular}[c]{@{}l@{}}Google data set\\ ($8350$ total time steps)\end{tabular}} & 37.86          & 554.97        \\ \hline
\end{tabular}
\end{center}
}
\end{table}

\begin{figure*}[t]
    \centering

     \begin{subfigure}{0.45\columnwidth}
        \centering
        \includegraphics[width=1\linewidth]{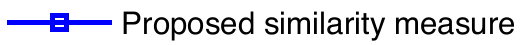}
    \end{subfigure}%
    \quad
     \begin{subfigure}{0.45\columnwidth}
        \centering
        \includegraphics[width=1\linewidth]{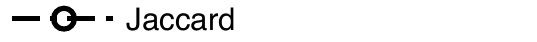}
    \end{subfigure}%
    \vspace{0.05in}

    \begin{subfigure}{0.33\columnwidth}
        \centering
        \includegraphics[width=1\linewidth]{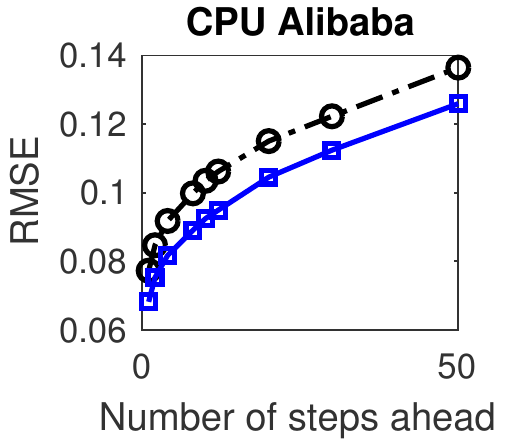}
    \end{subfigure}%
    \begin{subfigure}{0.33\columnwidth}
        \centering
        \includegraphics[width=1\linewidth]{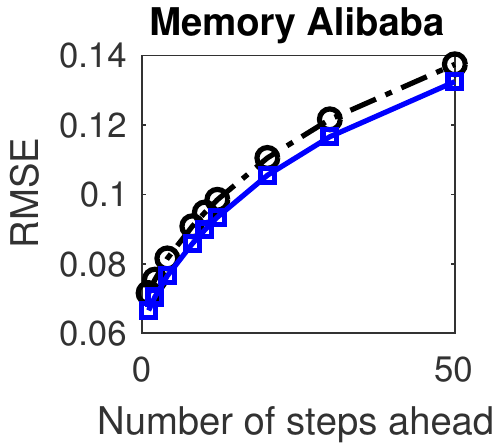}
    \end{subfigure}%
    \begin{subfigure}{0.33\columnwidth}
        \centering
        \includegraphics[width=1\linewidth]{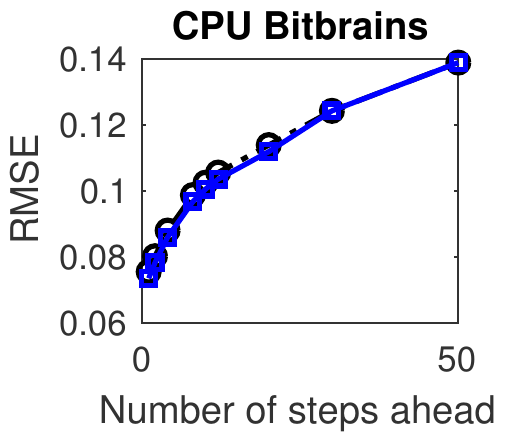}
    \end{subfigure}%
    \begin{subfigure}{0.33\columnwidth}
        \centering
        \includegraphics[width=1\linewidth]{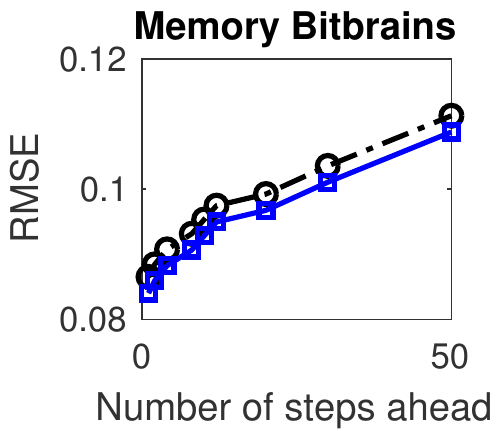}
    \end{subfigure}%
    \begin{subfigure}{0.33\columnwidth}
        \centering
        \includegraphics[width=1\linewidth]{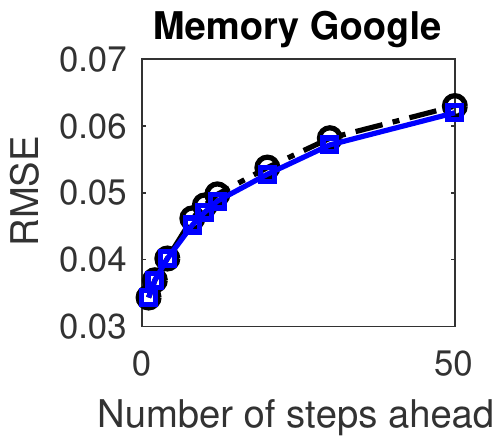}
    \end{subfigure}%
    \begin{subfigure}{0.33\columnwidth}
        \centering
        \includegraphics[width=1\linewidth]{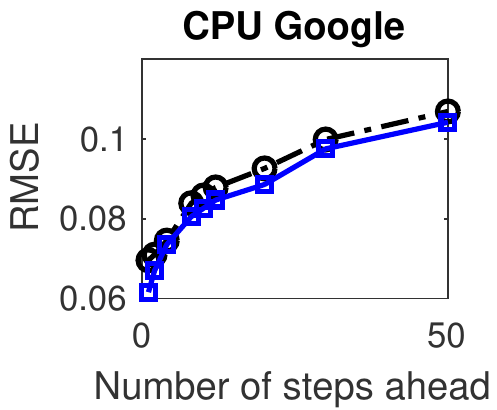}
    \end{subfigure}%

    \caption{Time-averaged RMSE with Jaccard Index and our proposed similarity measure.}
    \label{figure:jaccard}
    \vspace{-0.1in}
\end{figure*}

\begin{figure*}[t]
    \centering
   \begin{subfigure}{0.35\columnwidth}
    \centering
    \includegraphics[width=1\linewidth]{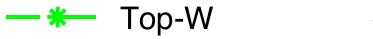}
    \end{subfigure}%
    \quad
    \begin{subfigure}{0.35\columnwidth}
    \centering
    \includegraphics[width=1\linewidth]{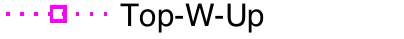}
    \end{subfigure}%
    \quad
     \begin{subfigure}{0.35\columnwidth}
        \centering
        \includegraphics[width=1\linewidth]{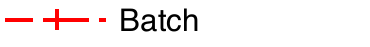}
    \end{subfigure}%
    \quad
     \begin{subfigure}{0.35\columnwidth}
        \centering
        \includegraphics[width=1\linewidth]{figure2/legend/legend-propap.pdf}
    \end{subfigure}%
      \quad
     \begin{subfigure}{0.35\columnwidth}
        \centering
        \includegraphics[width=1\linewidth]{figure2/legend/legend-min.pdf}
    \end{subfigure}%
    \vspace{0.05in}
    
    \begin{subfigure}{0.33\columnwidth}
        \centering
        \includegraphics[width=1\linewidth]{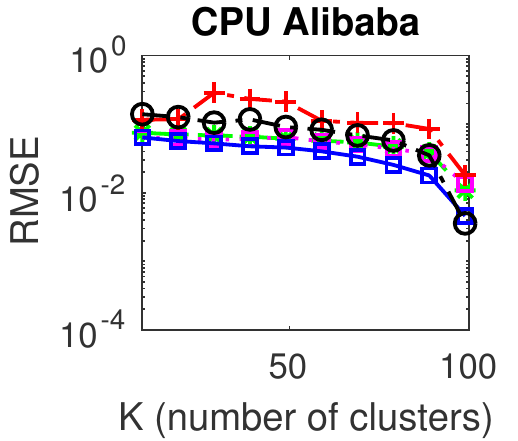}
    \end{subfigure}%
    \begin{subfigure}{0.33\columnwidth}
        \centering
        \includegraphics[width=1\linewidth]{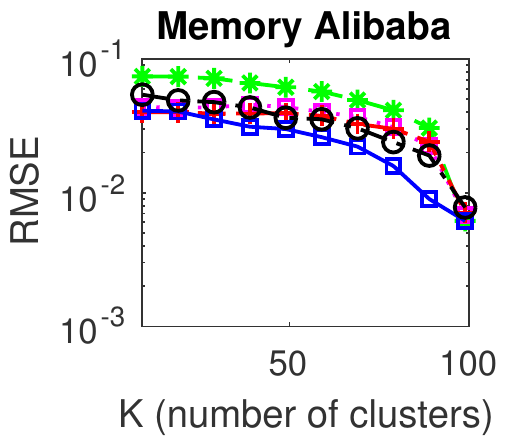}
    \end{subfigure}%
    \begin{subfigure}{0.33\columnwidth}
        \centering
        \includegraphics[width=1\linewidth]{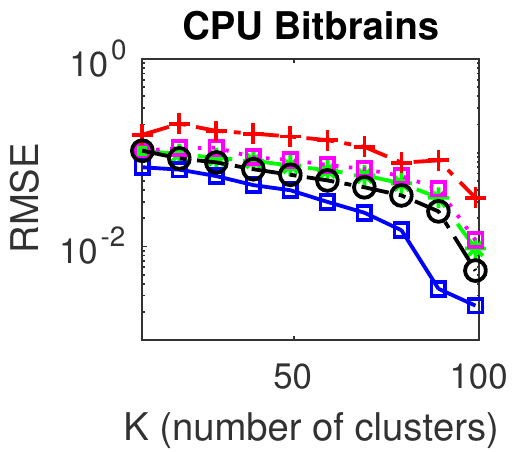}
    \end{subfigure}%
    \begin{subfigure}{0.33\columnwidth}
        \centering
        \includegraphics[width=1\linewidth]{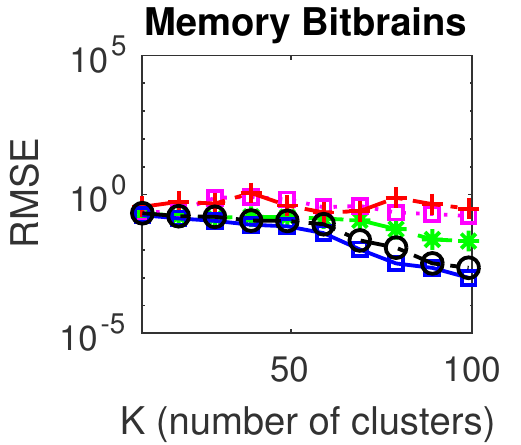}
    \end{subfigure}%
    \begin{subfigure}{0.33\columnwidth}
        \centering
        \includegraphics[width=1\linewidth]{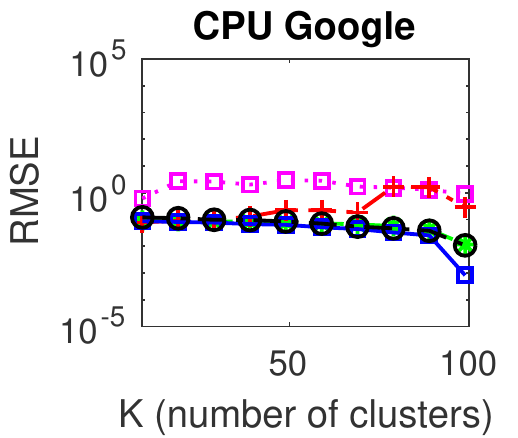}
    \end{subfigure}%
    \begin{subfigure}{0.33\columnwidth}
        \centering
        \includegraphics[width=1\linewidth]{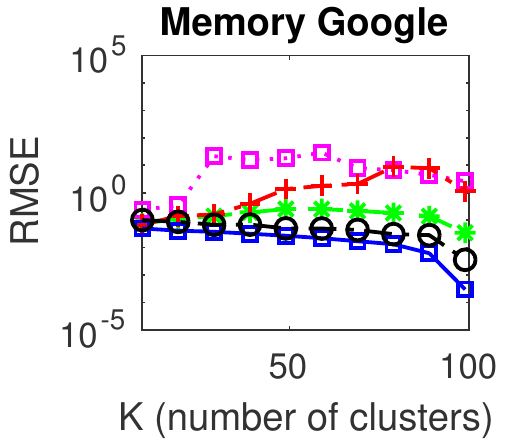}
    \end{subfigure}%
   
    \caption{RMSE for comparison with~\cite{silvestri2015online} with different number of clusters ($K$).}
    \label{figure:existing-work}
    \vspace{-0.15in}
\end{figure*}

\begin{table}[t]
\caption{RMSE with Different Values of $M$ and $M'$ for the Google dataset with CPU resource \vspace{-0.15in}}
{
\footnotesize
\begin{center}
$h=1$ \vspace{0.05in}

\begin{tabular}{|c|l|l|l|l|}
\hline
\textbf{} & ${M'=1}$ & ${M'=5}$ & ${M'=12}$ & ${M'=100}$ \\ \hline 
${M=1}$   &   0.055  & 0.068     &  0.071     &  0.106    \\ \hline
${M=5}$   &  0.058  &  0.068   &  0.068     &  0.098    \\ \hline
${M=12}$  &  0.059  &   0.048 &    0.046   &   0.050   \\ \hline
${M=100}$ & 0.065   &  0.089  &   0.047    &  0.055     \\ \hline
\end{tabular}
\vspace{0.1in}

$h=5$ \vspace{0.05in}

\begin{tabular}{|c|l|l|l|l|}
\hline
\textbf{} & ${M'=1}$ & ${M'=5}$ & ${M'=12}$ & ${M'=100}$ \\ \hline 
${M=1}$   & 0.088  &  0.073   & 0.076     &   0.108   \\ \hline
${M=5}$   & 0.105   &  0.081   &   0.074   & 0.099   \\ \hline
${M=12}$  & 0.117  & 0.079   &  0.076     &  0.097   \\ \hline
${M=100}$ & 0.091   & 0.0899  & 0.078      &   0.101    \\ \hline
\end{tabular}
\vspace{0.1in}

$h=10$ \vspace{0.05in}

\begin{tabular}{|c|l|l|l|l|}
\hline
\textbf{} & ${M'=1}$ & ${M'=5}$ & ${M'=12}$ & ${M'=100}$ \\ \hline 
${M=1}$   & 0.098     & 0.082      & 0.081       & 0.107      \\ \hline
${M=5}$   & 0.121     & 0.095      & 0.080       & 0.099     \\ \hline
${M=12}$  & 0.129      & 0.102      & 0.081      & 0.098      \\ \hline
${M=100}$ & 0.104      & 0.112      & 0.084       & 0.101      \\ \hline
\end{tabular}
\end{center}
}
\label{table:h1-cpu}
\end{table}

In the remaining of this subsection, we use the sample-and-hold method (with $K=3$) for forecasting and consider the impact of other aspects on the RMSE.

\subsubsection{Different Clustering Methods}\label{exp:diff-models}

We consider the different clustering methods as in Section~\ref{subsubsec:experiment-spatial-estimation-diff-clustering} combined with temporal forecasting. The RMSE results with different forecasting steps ($h$) are shown in Fig.~\ref{figure:forecast-sample}. We see that our proposed approach performs the best in almost all cases. For long-term forecasting with large $h$, the static clustering method often performs similar as our proposed approach, because when there are fluctuations, dynamic clustering may not perform as good as static clustering for long time periods. Note, however, that the static clustering baseline is an offline method which requires knowledge of the entire time series beforehand, thus it is not really applicable in practice.

\subsubsection{Different Values of $M$ and $M'$}\label{exp:M}
Table~\ref{table:h1-cpu} shows the RMSE with different values of $M$ and $M'$ on the Google dataset with CPU resource, where we recall that $M$ and $M'$ are the number of time steps to look back into history when computing the similarity measure and forecasted cluster (and per-node offset), respectively (see Sections~\ref{section:dynamic-cluster} and \ref{sec:temporalForecasting}).
We observe that the optimal choices of $M$ and $M'$ depend on the forecasting step $h$. Generally, $M=1$ is a reasonably good value for all cases. The optimal value of $M'$ tends to increase with $h$. This means that the farther-ahead we would like to forecast, the more we should look back into the history when determining the cluster membership and offset values of local nodes, which is intuitive because we need to rely more on long-term (stable) characteristics when forecasting farther ahead into the future. We choose $M'=5$ as default in Section~\ref{subsec:experiment-parameters} which is a relatively good value for different~$h$.

\subsubsection{Proposed Similarity Measure vs. Jaccard Index}\label{exp:jaccard}
As discussed in Section~\ref{section:dynamic-cluster}, the Jaccard index used in~\cite{greene2010tracking} is another possible similarity measure that one could use. In Fig.~\ref{figure:jaccard}, we compare the RMSE when using our proposed similarity measure and Jaccard index. Our proposed similarity measure gives a better or similar performance in all cases.

\subsection{Comparison to Gaussian-based Method in~\cite{silvestri2015online}}\label{exp:gaussian}

Finally, we modify our setup and compare our proposed approach with the \emph{Gaussian-based method} in~\cite{silvestri2015online}.

The method in \cite{silvestri2015online} includes separate training and testing phases, both set to $500$ time steps (which is the value chosen in~\cite{silvestri2015online}). During the training phase, the central node receives measurements from every node (i.e., $B=1$) and uses this information to select a subset of nodes ($K \ll N$) that will continue to send measurements during the testing phase. This subset of $K$ nodes is called \emph{monitors}.
During the testing phase, the central node receives measurements only from the selected nodes (which is equivalent to having a transmission frequency of $B=\frac{K}{N}$), and the measurements of the non-monitor nodes are inferred based on the measurements from the monitors. There is no temporal forecasting in this mechanism.

We adapt our proposed approach to the above setting with separate training and testing phases as follows. During training, we perform K-means clustering, where we group nodes into clusters based on their $500$ latest measurements (i.e., we perform K-means on $500$-dimensional vectors). This gives us $K$ clusters of nodes. We select one node in each cluster that has the smallest Euclidean distance from the centroid of this cluster. We consider this node as a monitor.
During testing, we only receive measurements from the monitors. The resource utilizations at all nodes that belong to the same cluster as the monitor are estimated as equal to the measurement of the monitor.
The minimum distance baseline in this setting is one that selects the $K$ monitors randomly, and the other nodes are assigned to clusters based on their Euclidean distances from the monitors, where each cluster contains one monitor.
Three algorithms that are proposed in \cite{silvestri2015online} are also considered as baselines: Top-W, Top-W-Update, and Batch Selection, which are based on Gaussian models. 

We only use $100$ randomly selected machines in this experiment, because the approaches in \cite{silvestri2015online} are too time-consuming to run on the entire dataset.
The results of RMSE defined on the estimation method described in this subsection above\footnote{Note that this RMSE definition is different from that in earlier parts of this paper.} are shown in Fig.~\ref{figure:existing-work} and the computational time of different approaches (on computer with Intel Core i7-6700 $3.4$~GHz CPU, $16$~GB memory) is shown in Table~\ref{table:time}. We see that our proposed approach provides the smallest RMSE, and it runs much faster than the three approaches (Top-W, Top-W-Update, and Batch Selection) from \cite{silvestri2015online}.
This observation is consistent with our discussions in Sections~\ref{section:related} and \ref{section:motivation-example} that Gaussian models do not work well in our setting.

\begin{table}[t]
\caption{Computation time (in seconds) for each approach and dataset (100 nodes) \vspace{-0.1in}}{
\footnotesize
\begin{center}
\begin{tabular}{llll}
\hline

                      & \textbf{CPU Alibaba} & \textbf{CPU Bitbrains} & \textbf{CPU Google} \\ \hline
\textbf{Proposed}     &       0.1401             &     0.16457       &         0.1370           \\ \hline
\textbf{Min.-distance} &        0.0231          &         0.0287           &      0.0238            \\ \hline
\textbf{Top-W}        &       0.5987            &       0.6134              &     0.6074              \\ \hline
\textbf{Top-W-Update}     &        29.3502          &       30.2132              &   27.4450                \\ \hline
\textbf{Batch Selection}        &         2.8197         &        2.7812              &  2.2934                \\ \hline
\end{tabular}
\end{center}
}
\label{table:time}
\end{table}

\section{Conclusion}\label{section:conclusion}

In this paper, we have proposed a novel mechanism for the efficient collection and forecasting of resource utilization at different machines in large-scale distributed systems.
The mechanism is a tight integration of algorithms for adaptive transmission, dynamic clustering, and temporal forecasting, with the goal of minimizing the RMSE of both spatial estimation and temporal forecasting.
Experiments on three real-world datasets show the effectiveness of our approach compared to baseline methods.
Future work can study the integration of our approach with resource allocation and other system management mechanisms.

\bibliographystyle{IEEEtran}
\bibliography{ref}

% Generated by IEEEtran.bst, version: 1.14 (2015/08/26)
\begin{thebibliography}{10}
\providecommand{\url}[1]{#1}
\csname url@samestyle\endcsname
\providecommand{\newblock}{\relax}
\providecommand{\bibinfo}[2]{#2}
\providecommand{\BIBentrySTDinterwordspacing}{\spaceskip=0pt\relax}
\providecommand{\BIBentryALTinterwordstretchfactor}{4}
\providecommand{\BIBentryALTinterwordspacing}{\spaceskip=\fontdimen2\font plus
\BIBentryALTinterwordstretchfactor\fontdimen3\font minus
  \fontdimen4\font\relax}
\providecommand{\BIBforeignlanguage}[2]{{%
\expandafter\ifx\csname l@#1\endcsname\relax
\typeout{** WARNING: IEEEtran.bst: No hyphenation pattern has been}%
\typeout{** loaded for the language `#1'. Using the pattern for}%
\typeout{** the default language instead.}%
\else
\language=\csname l@#1\endcsname
\fi
#2}}
\providecommand{\BIBdecl}{\relax}
\BIBdecl

\bibitem{CaiLessProvisioning2018}
\BIBentryALTinterwordspacing
B.~Cai, R.~Zhang, L.~Zhao, and K.~Li, ``Less provisioning: A fine-grained
  resource scaling engine for long-running services with tail latency
  guarantees,'' in \emph{Proceedings of the 47th International Conference on
  Parallel Processing}, ser. ICPP 2018.\hskip 1em plus 0.5em minus 0.4em\relax
  New York, NY, USA: ACM, 2018, pp. 30:1--30:11. [Online]. Available:
  \url{http://doi.acm.org/10.1145/3225058.3225113}
\BIBentrySTDinterwordspacing

\bibitem{Grechanik2016}
\BIBentryALTinterwordspacing
M.~Grechanik, Q.~Luo, D.~Poshyvanyk, and A.~Porter, ``Enhancing rules for cloud
  resource provisioning via learned software performance models,'' in
  \emph{Proceedings of the 7th ACM/SPEC on International Conference on
  Performance Engineering}, ser. ICPE '16.\hskip 1em plus 0.5em minus
  0.4em\relax New York, NY, USA: ACM, 2016, pp. 209--214. [Online]. Available:
  \url{http://doi.acm.org/10.1145/2851553.2851568}
\BIBentrySTDinterwordspacing

\bibitem{silvestri2015online}
S.~Silvestri, R.~Urgaonkar, M.~Zafer, and B.~J. Ko, ``An online method for
  minimizing network monitoring overhead,'' in \emph{Distributed Computing
  Systems (ICDCS), 2015 IEEE 35th International Conference on}.\hskip 1em plus
  0.5em minus 0.4em\relax IEEE, 2015, pp. 268--277.

\bibitem{Nikravesh2015}
\BIBentryALTinterwordspacing
A.~Y. Nikravesh, S.~A. Ajila, and C.-H. Lung, ``Towards an autonomic
  auto-scaling prediction system for cloud resource provisioning,'' in
  \emph{Proceedings of the 10th International Symposium on Software Engineering
  for Adaptive and Self-Managing Systems}, ser. SEAMS '15.\hskip 1em plus 0.5em
  minus 0.4em\relax Piscataway, NJ, USA: IEEE Press, 2015, pp. 35--45.
  [Online]. Available: \url{http://dl.acm.org/citation.cfm?id=2821357.2821365}
\BIBentrySTDinterwordspacing

\bibitem{Shen2018}
H.~Shen and L.~Chen, ``Resource demand misalignment: An important factor to
  consider for reducing resource over-provisioning in cloud datacenters,''
  \emph{IEEE/ACM Transactions on Networking}, vol.~26, no.~3, pp. 1207--1221,
  June 2018.

\bibitem{coluccia2011lossy}
G.~Coluccia, E.~Magli, A.~Roumy, and V.~Toto-Zarasoa, ``Lossy compression of
  distributed sparse sources: a practical scheme,'' in \emph{Signal Processing
  Conference, 2011 19th European}.\hskip 1em plus 0.5em minus 0.4em\relax IEEE,
  2011, pp. 422--426.

\bibitem{barcelo2012enhanced}
J.~E. Barcel{\'o}-Llad{\'o}, A.~M. P{\'e}rez, and G.~Seco-Granados, ``Enhanced
  correlation estimators for distributed source coding in large wireless sensor
  networks,'' \emph{IEEE Sensors Journal}, vol.~12, no.~9, pp. 2799--2806,
  2012.

\bibitem{anagnostopoulos2014advanced}
C.~Anagnostopoulos and S.~Hadjiefthymiades, ``Advanced principal
  component-based compression schemes for wireless sensor networks,'' \emph{ACM
  Transactions on Sensor Networks (TOSN)}, vol.~11, no.~1, p.~7, 2014.

\bibitem{li2018compressed}
Y.~Li and Y.~Liang, ``Compressed sensing in multi-hop large-scale wireless
  sensor networks based on routing topology tomography,'' \emph{IEEE Access},
  vol.~6, pp. 27\,637--27\,650, 2018.

\bibitem{leinonen2014compressed}
M.~Leinonen, M.~Codreanu, and M.~Juntti, ``Compressed acquisition and
  progressive reconstruction of multi-dimensional correlated data in wireless
  sensor networks,'' in \emph{Acoustics, Speech and Signal Processing (ICASSP),
  2014 IEEE International Conference on}.\hskip 1em plus 0.5em minus
  0.4em\relax IEEE, 2014, pp. 6449--6453.

\bibitem{zhang2016near}
Y.~Zhang, T.~N. Hoang, K.~H. Low, and M.~S. Kankanhalli, ``Near-optimal active
  learning of multi-output gaussian processes.'' in \emph{AAAI}, 2016, pp.
  2351--2357.

\bibitem{krause2008near}
A.~Krause, A.~Singh, and C.~Guestrin, ``Near-optimal sensor placements in
  gaussian processes: Theory, efficient algorithms and empirical studies,''
  \emph{Journal of Machine Learning Research}, vol.~9, no. Feb, pp. 235--284,
  2008.

\bibitem{liu2005energy}
C.~Liu, K.~Wu, and M.~Tsao, ``Energy efficient information collection with the
  arima model in wireless sensor networks,'' in \emph{Global Telecommunications
  Conference, 2005. GLOBECOM'05. IEEE}, vol.~5.\hskip 1em plus 0.5em minus
  0.4em\relax IEEE, 2005, pp. 5--pp.

\bibitem{law2009energy}
Y.~W. Law, S.~Chatterjea, J.~Jin, T.~Hanselmann, and M.~Palaniswami,
  ``Energy-efficient data acquisition by adaptive sampling for wireless sensor
  networks,'' in \emph{Proceedings of the 2009 International Conference on
  Wireless Communications and Mobile Computing: Connecting the World
  Wirelessly}.\hskip 1em plus 0.5em minus 0.4em\relax ACM, 2009, pp.
  1146--1151.

\bibitem{harb2016adaptive}
H.~Harb, A.~Makhoul, A.~Jaber, R.~Tawil, and O.~Bazzi, ``Adaptive data
  collection approach based on sets similarity function for saving energy in
  periodic sensor networks,'' \emph{International Journal of Information
  Technology and Management}, vol.~15, no.~4, pp. 346--363, 2016.

\bibitem{chatterjea2008adaptive}
S.~Chatterjea and P.~Havinga, ``An adaptive and autonomous sensor sampling
  frequency control scheme for energy-efficient data acquisition in wireless
  sensor networks,'' in \emph{International Conference on Distributed Computing
  in Sensor Systems}.\hskip 1em plus 0.5em minus 0.4em\relax Springer, 2008,
  pp. 60--78.

\bibitem{idrees2017distributed}
A.~K. Idrees and A.~K.~M. Al-Qurabat, ``Distributed adaptive data collection
  protocol for improving lifetime in periodic sensor networks.'' \emph{IAENG
  International Journal of Computer Science}, vol.~44, no.~3, 2017.

\bibitem{chakrabarti2006evolutionary}
D.~Chakrabarti, R.~Kumar, and A.~Tomkins, ``Evolutionary clustering,'' in
  \emph{Proceedings of the 12th ACM SIGKDD international conference on
  Knowledge discovery and data mining}.\hskip 1em plus 0.5em minus 0.4em\relax
  ACM, 2006, pp. 554--560.

\bibitem{xu2014adaptive}
K.~S. Xu, M.~Kliger, and A.~O. Hero~III, ``Adaptive evolutionary clustering,''
  \emph{Data Mining and Knowledge Discovery}, vol.~28, no.~2, pp. 304--336,
  2014.

\bibitem{greene2010tracking}
D.~Greene, D.~Doyle, and P.~Cunningham, ``Tracking the evolution of communities
  in dynamic social networks,'' in \emph{Advances in social networks analysis
  and mining (ASONAM), 2010 international conference on}.\hskip 1em plus 0.5em
  minus 0.4em\relax IEEE, 2010, pp. 176--183.

\bibitem{yang2011detecting}
T.~Yang, Y.~Chi, S.~Zhu, Y.~Gong, and R.~Jin, ``Detecting communities and their
  evolutions in dynamic social networks—a bayesian approach,'' \emph{Machine
  learning}, vol.~82, no.~2, pp. 157--189, 2011.

\bibitem{ma2006evolutionary}
P.~C. Ma, K.~C. Chan, X.~Yao, and D.~K. Chiu, ``An evolutionary clustering
  algorithm for gene expression microarray data analysis,'' \emph{IEEE
  Transactions on Evolutionary Computation}, vol.~10, no.~3, pp. 296--314,
  2006.

\bibitem{rana2014evolutionary}
C.~Rana and S.~K. Jain, ``An evolutionary clustering algorithm based on
  temporal features for dynamic recommender systems,'' \emph{Swarm and
  Evolutionary Computation}, vol.~14, pp. 21--30, 2014.

\bibitem{bodik2004intel}
P.~Bodik, W.~Hong, C.~Guestrin, S.~Madden, M.~Paskin, and R.~Thibaux, ``Intel
  lab data,'' \emph{Online dataset}, 2004.

\bibitem{reiss2011google}
C.~Reiss, J.~Wilkes, and J.~L. Hellerstein, ``Google cluster-usage traces:
  format+ schema,'' \emph{Google Inc., White Paper}, pp. 1--14, 2011.

\bibitem{Aloise2009}
D.~Aloise, A.~Deshpande, P.~Hansen, and P.~Popat, ``Np-hardness of euclidean
  sum-of-squares clustering,'' \emph{Machine Learning}, vol.~75, no.~2, pp.
  245--248, May 2009.

\bibitem{neely2010stochastic}
M.~J. Neely, ``Stochastic network optimization with application to
  communication and queueing systems,'' \emph{Synthesis Lectures on
  Communication Networks}, vol.~3, no.~1, pp. 1--211, 2010.

\bibitem{KMeansAlg}
J.~A. Hartigan and M.~A. Wong, ``Algorithm as 136: A k-means clustering
  algorithm,'' \emph{Journal of the Royal Statistical Society. Series C
  (Applied Statistics)}, vol.~28, no.~1, pp. 100--108, 1979.

\bibitem{kuhn1955hungarian}
H.~W. Kuhn, ``The hungarian method for the assignment problem,'' \emph{Naval
  research logistics quarterly}, vol.~2, no. 1-2, pp. 83--97, 1955.

\bibitem{WARRENLIAO20051857}
\BIBentryALTinterwordspacing
T.~W. Liao, ``Clustering of time series data—a survey,'' \emph{Pattern
  Recognition}, vol.~38, no.~11, pp. 1857 -- 1874, 2005. [Online]. Available:
  \url{http://www.sciencedirect.com/science/article/pii/S0031320305001305}
\BIBentrySTDinterwordspacing

\bibitem{box1974some}
G.~E. Box, G.~M. Jenkins, and J.~F. MacGregor, ``Some recent advances in
  forecasting and control,'' \emph{Applied Statistics}, pp. 158--179, 1974.

\bibitem{hochreiter1997long}
S.~Hochreiter and J.~Schmidhuber, ``Long short-term memory,'' \emph{Neural
  computation}, vol.~9, no.~8, pp. 1735--1780, 1997.

\bibitem{alibaba}
``{Alibaba trace},'' \url{https://github.com/alibaba/clusterdata/tree/v2018},
  2018.

\bibitem{shen2014workload}
S.~Shen, V.~Van~Beek, and A.~Iosup, ``Workload characterization of cloud
  datacenter of bitbrains,'' \emph{TU Delft, Tech. Rep. PDS-2014-001}, 2014.

\bibitem{burnham2003model}
K.~P. Burnham and D.~R. Anderson, \emph{Model selection and multimodel
  inference: a practical information-theoretic approach}.\hskip 1em plus 0.5em
  minus 0.4em\relax Springer Science \& Business Media, 2003.

\end{thebibliography}

\end{document}